\pdfoutput=1

\documentclass[12pt,english,pra,reprint,twocolumn,en]{revtex4-1}
\usepackage[latin9]{inputenc}
\usepackage{xcolor}
\usepackage{babel}
\usepackage{bm}
\usepackage{amsmath}
\usepackage{amssymb}
\usepackage{graphicx}
\usepackage[unicode=true,pdfusetitle,
 bookmarks=true,bookmarksnumbered=false,bookmarksopen=false,
 breaklinks=false,pdfborder={0 0 0},pdfborderstyle={},backref=false,colorlinks=true]
 {hyperref}
\hypersetup{
 linkcolor=blue,citecolor=magenta,urlcolor=cyan}

\makeatletter
\usepackage{babel}
\usepackage{babel}
\pdfoutput=1
\definecolor{green}{RGB}{0, 180, 0}
\definecolor{cyan}{RGB}{0, 180, 180}
\definecolor{yellow}{RGB}{211,211,0}

\makeatother

\begin{document}
\title{Quantization of graphene plasmons}
\author{Beatriz A. Ferreira$^{1,2}$, B. Amorim$^{1,3}$, A. J. Chaves$^{4}$,
and N. M. R. Peres$^{1,2}$}
\email[Contact e-mail:]{peres@fisica.uminho.pt}

\address{$^{1}$Department of Physics, Center of Physics, and QuantaLab, University
of Minho, Campus of Gualtar, 4710-057, Braga, Portugal}
\address{$^{2}$International Iberian Nanotechnology Laboratory (INL), Av. Mestre
José Veiga, 4715-330 Braga, Portugal}
\address{$^{3}$CeFEMA, Instituto Superior Técnico, Universidade de Lisboa,
Av. Rovisco Pais, 1049-001 Lisboa, Portugal}
\address{$^{4}$Department of Physics, Instituto Tecnológico de Aeronáutica,
DCTA, 12228-900 São José dos Campos, Brazil}
\begin{abstract}
In this article we perform the quantization of graphene plasmons using
both a macroscopic approach based on the classical average electromagnetic
energy and a quantum hydrodynamic model, in which graphene charge
carriers are modeled as a charged fluid. Both models allow to take
into account the dispersion of graphenes optical response, with the
hydrodynamic model also allowing for the inclusion of non-local effects.
Using both methods, the electromagnetic field mode-functions, and
the respective frequencies, are determined for two different graphene
structures. we show how to quantize graphene plasmons, considering
that graphene is a dispersive medium, and taking into account both
local and nonlocal descriptions. It is found that the dispersion of
graphene's optical response leads to a non-trivial normalization condition
for the mode-functions. The obtained mode-functions are then used
to calculate the decay of an emitter, represented by a dipole, via
the excitation of graphene surface plasmon-polaritons. The obtained
results are compared with the total spontaneous decay rate of the
emitter and a near perfect match is found in the relevant spectral
range. It is found that non-local effects in graphene's conductivity,
become relevant for the emission rate for small Fermi energies and
small distances between the dipole and the graphene sheet.
\end{abstract}
\maketitle

\section{Introduction}

In many cases, light-matter interaction can be understood in a semi-classical
picture, where matter is quantized and the electromagnetic field (EM)
is treated classically. This semi-classical approach holds when the
number of photons in the field is large or the light source is coherent.
On the other hand, in order to understand the properties of a small
number of photons the quantization of the EM field is required. Typical
phenomena where the quantization of the EM field is necessary involve
entanglement, squeezed light, cavity electrodynamics, interaction
of photons with nano-mechanical resonators, and near-field radiative
effects \citep{Agarwal2013}.

In near-field radiative effects, plasmonics emerges as a promising
candidate to observe quantum effects of the electromagnetic radiation,
an example being the Hong-Ou-Mandel interference of plasmons \citep{Heeres:2013aa}.
Many other quantum effects in plasmonics exist, such as the survival
of entanglement, particle-wave duality, quantum size effects due to
reduced dimensions of metallic nanostructures, quantum tunneling of
plasmons (which are simultaneously light and matter), and coupling
of quantum emitters to surface plasmons \citep{Jacob2012,Tame2013,ModernPlasm,Fakonas:2015aa,Maier2016,Dheur:2016aa,Bozhevolnyi2017,Xu:2018aa,Mortensen2018}.

The exploration of quantum effects in plasmonics in unusual spectral
ranges, such as the THz and the mid-IR, has been deterrent by the
poor plasmonic response of noble metals in these regions of the electromagnetic
spectrum. However, with the emergence of graphene plasmonics \citep{Ju2011,Fei2011}
the possibility of exploring quantum effects in these yet unexplored
spectral regions is a possible prospect. Despite the fact that, at
the time of writing, quantum effects involving graphene plasmons remain
illusive,\textcolor{teal}{{} }the fact that graphene plasmons are characterized
by low losses \citep{Fei2012,Chen2012,Woessner2015} boosts the above
hope. In addition, graphene plasmons screened by a nearby metal (also
called screened or acoustic plasmons) can be confined down to the
atomic limit \citep{Iranzo2018}, which certainly opens the prospects
of finding rich grounds for quantum plasmonics and nonlocal effects
\citep{Lundeberg:2017aa,Dias:2018aa}. Indeed, the idea of developing
quantum optics with plasmons has already a long history \citep{Chang2006}
and quantization of localized plasmons in metallic nanoparticles was
recently performed \citep{Alpeggiani:2014aa,Bordo:2019aa}.

\begin{figure}
\centering{}\includegraphics[width=8cm]{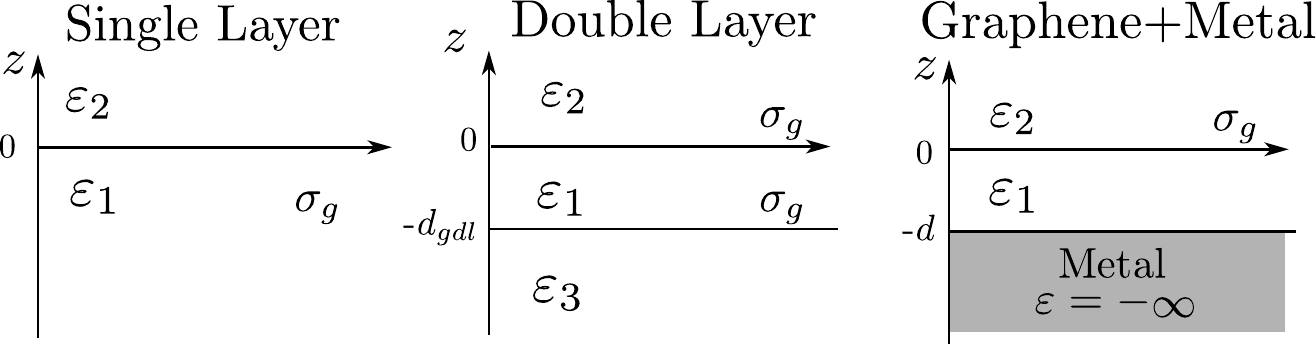}\caption{\label{fig:three_systems}Schematic representation of the three systems
considered in this paper: a single graphene layer (left), and graphene
double layer (center), and a graphene sheet near a perfect metal.
The quantities $\epsilon_{n}$ refer to the dielectric permittivity
of the medium $n$ and $\sigma_{g}$ refers to the optical conductivity
of graphene.}
\end{figure}

The development of quantum theory of the electromagnetic field in
the presence of dieletric media has a long history and several approaches
have been developed \citep{Hopfield1958,Alekseev1966,Agarwal1975_I,Huttner1992,Milonni1995,Gruner1995,Gruner1996,Matloob1995,Dung1998,Philbin2010,Tame2015,Allameh2016,Sha2018}.
These methods are typically based either on the quantization of the
macroscopic classic energy\citep{Milonni1995}, on the classical dyadic
Green's function for the electric field\citep{Gruner1995} or on the
diagonalization of Hopfield-type Hamiltonians\citep{Huttner1992}.
The quantization of evanescent EM waves \citep{Carniglia1972,Carniglia:1972aa}
and of the EM field in the vicinity of a metal \citep{Grossel:1977aa}
have also been considered in the past.

In this paper, we perform the quantization of graphene plasmons, obtaining
the plasmonic electromagnetic field mode-functions and, importantly,
their normalization, when losses are neglected. These mode-functions
are then used to study the interaction of graphene plasmons with nearby
quantum emitters and determine, in a very intuitive way, using Fermi's
golden rule the spontaneous decay rate of the emitter due to plasmon
emission. Thee quantization of graphene plasmons is performed in two
ways: (i) A macroscopic approach, which starts from the classical
time-averaged energy of the electromagnetic field in a dielectric
\citep{Landau_Electrodynamics,Alekseev1966,Agu1979,Milonni1995}.
This method allows for the inclusion of dispersion in the optical
response of graphene. (ii) A hydrodynamic approach, where graphene
charge carriers are described in terms of an electronic fluid, which
couples to the eletromagnetic field \citep{Chaves2017,Fetter1973}.
The hydrodynamic approach allows not only for the inclusion of dispersion,
but also for the inclusion of non-local effects in the optical response
of graphene.

The paper is organized as follows: in Section~\ref{sec:macroscopic_quantization},
we present the general macroscopic approach for the quantization of
the eletromagnetic field in dispersive, lossless media and a normalization
condition for the mode-functions is determined. In Section~\ref{sec:hydrodynamic_quantum_model},
we present the quantum hydrodynamic model or graphene. We will see
that when non-local effects are neglected, the result of the macroscopic
approach is recovered. In Section~\ref{sec:plasmon_dispersion},
the plasmon dispersion relations, mode-functions and mode-function
normalization for a single graphene layer and for a graphene-metal
structures are obtained. In Section~\ref{sec:Plasmon_emission_rate},
we use the quantized plasmon fields to compute the decay rate of a
quantum emitter due to the spontaneous emission of plasmons. The plasmon
emission rate is compared with the total decay rate of the emitter,
which is obtained from the complete dyadic Green's function for the
electric field. The role of non-local response of graphene is analyzed.
Finally, we conclude with Section~\ref{sec:Conclusions}, commenting
the obtained results and discussing future research paths. A set of
appendices detailing the calculations is also presented.

\section{Macroscopic quantization of the plasmon electromagnetic field\label{sec:macroscopic_quantization}}

In this section, we will describe how the plasmon fields can be quantized
using the macroscopic classical energy of the electromagnetic field
in a dispersive, lossless dielectric medium, a method first used in
Refs.~\citep{Alekseev1966,Agu1979,Milonni1995}. For electric and
magnetic fields with a harmonic time dependence,
\begin{align}
\mathbf{E}(\mathbf{r},t) & =\mathbf{E}_{\omega}(\mathbf{r})e^{-i\omega t}+\text{c.c.}\\
\mathbf{B}(\mathbf{r},t) & =\mathbf{B}_{\omega}(\mathbf{r})e^{-i\omega t}+\text{c.c.}
\end{align}
close to a central frequency $\omega$, the time-averaged classical
electromagnetic energy in the presence of a dispersive, lossless dieletric
is given by \citep{Landau_Electrodynamics,Ruppin_2002}
\begin{multline}
U_{\text{EM}}(\omega)=\int d^{3}\mathbf{r}\left(\epsilon_{0}\mathbf{E}_{\omega}^{*}(\mathbf{r})\cdot\frac{\partial}{\partial\omega}\left[\omega\bar{\epsilon}_{r}(\mathbf{r},\omega)\right]\cdot\mathbf{E}_{\omega}(\mathbf{r})\right.\\
\left.+\frac{1}{\mu_{0}}\left|\mathbf{B}_{\omega}\right|^{2}\right),
\end{multline}
where $\bar{\epsilon}_{r}(\mathbf{r},\omega)$ is the relative dielectric
tensor of the medium, $\epsilon_{0}$ and $\mu_{0}$ are, respectively,
vacuum permittivity and permeability. The idea of this method is to
take the above equation as the quantum mechanical energy of a EM field
eigenmode with frequency $\omega$. We will work in the Weyl gauge,
in which the scalar potential is set to zero $\phi=0$, such that
the electric and magnetic fields are obtained only from the vector
potential $\mathbf{A}$ as 
\begin{align}
\mathbf{E}(\mathbf{r},t) & =-\frac{\partial\mathbf{A}(\mathbf{r},t)}{\partial t},\label{eq:E_Weyl}\\
\mathbf{B}(\mathbf{r},t) & =\nabla\times\mathbf{A}(\mathbf{r},t).\label{eq:B_Weyl}
\end{align}
The vector potential is then expanded in modes as
\begin{equation}
\mathbf{A}(\mathbf{r},t)=\sum_{\lambda}\alpha_{\lambda}e^{-i\omega_{\lambda}t}\mathbf{A}_{\lambda}(\mathbf{r})+\text{c.c.},\label{eq:vector_potential_expansion}
\end{equation}
where $\alpha_{\lambda}$ are amplitudes for the mode $\lambda$,
with mode-function $\mathbf{A}_{\lambda}(\mathbf{r})$ and corresponding
frequency $\omega_{\lambda}$. The mode-functions and frequencies
are determined by solving the non-linear eigenvalue problem
\begin{equation}
\nabla\times\nabla\times\mathbf{A}_{\lambda}(\mathbf{r})=\frac{\omega_{\lambda}^{2}}{c^{2}}\bar{\epsilon}_{r}(\mathbf{r},\omega_{\lambda})\cdot\mathbf{A}_{\lambda}(\mathbf{r}),\label{eq:mode-equation}
\end{equation}
with $c$ vacuum's speed of light, which is just Ampère's law in the
dielectric medium for the mode-function $\mathbf{A}_{\lambda}(\mathbf{r})$.
Next, we assume that the total time-averaged energy for the vector
potential Eq.~\eqref{eq:vector_potential_expansion} is given by
\begin{equation}
U_{\text{EM}}=\sum_{\lambda}U_{\text{EM}}(\omega_{\lambda})\left|\alpha_{\lambda}\right|^{2}.\label{eq:time-averaged_energy}
\end{equation}
The quantization of the theory is achieved by promoting the amplitudes
$\alpha_{\lambda}$ to quantum mechanical operators
\begin{align}
\alpha_{\lambda} & \rightarrow\sqrt{\frac{\hbar}{2L_{\lambda}\epsilon_{0}\omega_{\lambda}}}\hat{a}_{\lambda},\label{eq:quantization_a}\\
\alpha_{\lambda}^{\dagger} & \rightarrow\sqrt{\frac{\hbar}{2L_{\lambda}\epsilon_{0}\omega_{\lambda}}}\hat{a}_{\lambda}^{\dagger},\label{eq:quantization_ad}
\end{align}
where $\hat{a}_{\lambda}$ ($\hat{a}_{\lambda}^{\dagger}$) are bosonic
annihilation (creation) operators, which obey the usual equal-time
commutation relations
\begin{equation}
\left[\hat{a}_{\lambda},\hat{a}_{\lambda^{\prime}}^{\dagger}\right]=\delta_{\lambda,\lambda^{\prime}},\label{eq:commutator_a_a_dagger}
\end{equation}
and $L_{\lambda}$ is a mode-length which is determined by demanding
that the quantum mechanical Hamiltonian which is obtained from Eq.~\eqref{eq:time-averaged_energy}
by performing the replacement of Eqs\@.~\eqref{eq:quantization_a}
and \eqref{eq:quantization_ad},
\begin{equation}
\hat{H}=\sum_{\lambda}\frac{\hbar U(\omega_{\lambda})}{4L_{\lambda}\epsilon_{0}\omega_{\lambda}}\left(\hat{a}_{\lambda}^{\dagger}\hat{a}_{\lambda}+\hat{a}_{\lambda}\hat{a}_{\lambda}^{\dagger}\right),
\end{equation}
coincides with the Hamiltonian for a collection of quantum harmonic
oscillators
\begin{equation}
\hat{H}=\frac{1}{2}\sum_{\lambda}\hbar\omega_{\lambda}\left(\hat{a}_{\lambda}^{\dagger}\hat{a}_{\lambda}+\hat{a}_{\lambda}\hat{a}_{\lambda}^{\dagger}\right).
\end{equation}
Imposing this condition, we have that the mode-length is given by
$L_{\lambda}=U(\omega_{\lambda})/\left(2\epsilon_{0}\omega_{\lambda}^{2}\right)$.
Using the mode-function equation \eqref{eq:mode-equation} to write
\begin{multline}
\int d^{3}\mathbf{r}\left|\nabla\times\mathbf{A}_{\lambda}(\mathbf{r})\right|^{2}=\int d^{3}\mathbf{r}\mathbf{A}_{\lambda}^{*}(\mathbf{r})\cdot\nabla\times\nabla\times\mathbf{A}_{\lambda}(\mathbf{r})\\
=\frac{\omega_{\lambda}^{2}}{c^{2}}\int d^{3}\mathbf{r}\mathbf{A}_{\lambda}^{*}(\mathbf{r})\cdot\bar{\epsilon}_{r}(\mathbf{r},\omega_{\lambda})\cdot\mathbf{A}_{\lambda}(\mathbf{r}),
\end{multline}
The mode-length can be written as
\begin{equation}
L_{\lambda}=\int d^{3}\mathbf{r}\mathbf{A}_{\lambda}^{*}(\mathbf{r})\cdot\left(\bar{\epsilon}_{r}(\mathbf{r},\omega_{\lambda})+\frac{\omega_{\lambda}}{2}\frac{\partial}{\partial\omega}\bar{\epsilon}_{r}(\mathbf{r},\omega_{\lambda})\right)\cdot\mathbf{A}_{\lambda}(\mathbf{r}).\label{eq:mode_length}
\end{equation}
Notice that in the absence of dispersion, the second term vanishes
and $L_{\lambda}$ reduces to the usual norm in the presence of a
position dependent dielectric tensor $\bar{\epsilon}_{r}(\mathbf{r},\omega_{\lambda})$.

Although this phenomenological method appears to be unjustified, it
has actually been shown to be correct for the case in when the dielectric
is modeled by a Lorentz oscillator \citep{Agu1979}. We will see in
the next section, that Eq.~\ref{eq:mode_length} remains valid within
a quantum hydrodynamic model of graphene, even when non-local effects
are included in the optical response. As a matter of fact Eq.~\eqref{eq:mode_length}
for the mode-length remains valid for any linear optical medium (including
effects of dispersion, non-locality, inhomogeneity and anisotropy)
as long as losses can be neglected \citep{Amorim2019}.

\section{plasmon quantization within a quantum hydrodynamic model\label{sec:hydrodynamic_quantum_model}}

In this section, we will perform the quantization of graphene surface
plasmons employing a hydrodynamic model. The hydrodynamic model treats
both the electron gas and the electromagnetic field using a classical
picture and provides a simple and elegant way of including non-local
effects \citep{christensen2017classical}. Non-local effects are taken
into account by including a pressure term in the Boltzmann equation,
that arises due to Pauli's exclusion principle. A detailed derivation
of the hydrodynamic model for graphene can be found in \citep{Chaves2017,Fetter1973}.
To illustrate the method, we choose the simple case of a single graphene
sheet located at the plane $z=0$, embedded by a static dielectric
medium with relative dielectric constant $\bar{\epsilon}_{d}(\mathbf{r})$.

\subsection{Classical hydrodynamic Lagrangean and Hamiltonian}

The classical Lagrangean density for the hydrodynamic model of graphene
can be written as: 
\begin{equation}
\mathcal{L}=\mathcal{L}_{\text{EM}}+\mathcal{L}_{\text{2D hyd}},\label{eq:lagrangean_EM+hydro}
\end{equation}
where $\mathcal{L}_{\text{EM}}$ is the Lagrangean density of the
electromagnetic field and $\mathcal{L}_{\text{2D hyd}}$ describes
the electronic fluid of graphene and its coupling to the electromagnetic
field. Using once again the Weyl gauge, $\mathcal{L}_{\text{EM}}$
is given by
\begin{equation}
\mathcal{L}_{\text{EM}}=\frac{\epsilon_{0}}{2}\left(\partial_{t}\mathbf{A}\right)\cdot\bar{\epsilon}_{d}\cdot\left(\partial_{t}\mathbf{A}\right)-\frac{1}{2\mu_{0}}\left(\nabla\times\mathbf{A}\right)^{2},
\end{equation}
where $\bar{\epsilon}_{d}(\mathbf{r})$ is allowed to be position
dependent, but is frequency independent. Within the hydrodynamic model,
the electronic fluid of the graphene layer is described by the fluctuation,
$n$, of the density around the equilibrium density, $n_{0}$, and
the displacement vector $\bm{\upsilon}=\left(\upsilon_{x},\upsilon_{y},0\right)$,
which should not be confused with the velocity field. In the Weyl
gauge, $\mathcal{L}_{\text{2D fluid}}$ is written as\citep{Fetter1973,Chaves2017}
\begin{multline}
\mathcal{L}_{\text{2D hyd}}=\delta(z)\Biggl(\frac{1}{2}n_{0}m\left(\partial_{t}\bm{\upsilon}\right)^{2}+m\beta^{2}n\nabla\cdot\bm{\upsilon}\\
\left.+\frac{m\beta^{2}}{2n_{0}}n^{2}-n_{0}e\partial_{t}\bm{\upsilon}\cdot\mathbf{A}\right),\label{eq:lagrangean_hydro}
\end{multline}
where the $\delta$-function $\delta(z)$ restricts the electronic
fluid to the $z=0$ 2D plane, $m$ is the Drude mass and $\beta$
appears from the relation between the degeneracy pressure and the
electronic density and depends on the band dispersion for the carriers
(see Ref.~\citep{Chaves2017}). In the approximation of the linear
dispersion for graphene electrons, the hydrodynamic parameters are
given by \citep{Chaves2017}: $n_{0}=k_{F}^{2}/\pi$, $m=\hbar k_{F}/v_{F}$
and $\beta^{2}=v_{F}^{2}/2$, where $k_{F}$ is the Fermi wavenumber
and $v_{F}$ the Fermi velocity of graphene. Equation~\eqref{eq:lagrangean_hydro}
is the 2D equivalent Lagrangian for the hydrodynamic model presented
in \citep{Nakamura1983}.

Using the Euler-Lagrange equations for Eq.~\eqref{eq:lagrangean_hydro}
with respect to $\mathbf{A}$, we obtain
\begin{equation}
\nabla\times\mathbf{B}=\frac{1}{c^{2}}\bar{\epsilon}_{d}\partial_{t}\mathbf{E}-\mu_{0}n_{0}e\partial_{t}\boldsymbol{\upsilon}\delta(z),\label{eq:Ampere}
\end{equation}
which is nothing more than Ampère's law in the presence of a surface
current given by
\begin{equation}
\mathbf{J}_{s}^{\text{hyd}}=-en_{0}\partial_{t}\boldsymbol{\upsilon}.\label{eq:hydro_current}
\end{equation}
Using the Euler-Lagrange equations for Eq.~\eqref{eq:lagrangean_hydro}
with respect to the fluid variables $n$ and $\boldsymbol{\upsilon}$,
we obtain the continuity and Newton's second law with a diffusion
term, which read respectively
\begin{align}
n_{0}\nabla\cdot\boldsymbol{\upsilon}+n & =0,\label{eq:continuity_2D}\\
n_{0}m\partial_{t}^{2}\boldsymbol{\upsilon}+m\beta^{2}\nabla n & =-en_{0}\mathbf{E}(z=0),\label{eq:second_law}
\end{align}
from which we recognize the fluid electronic surface density
\begin{equation}
\rho_{s}^{\text{hyd}}=-en.
\end{equation}
Equations~\eqref{eq:Ampere}-\eqref{eq:second_law} correspond to
the linearized hydrodynamic model for graphene \citep{Chaves2017}
(see also \citep{Lucas2018}).

Notice that Eq.~\eqref{eq:continuity_2D} has no dynamics. Therefore,
we can use it to eliminate the field $n$, thus obtaining a new Lagrangean
density
\begin{equation}
\mathcal{L}^{\prime}=\mathcal{L}_{\text{EM}}+\mathcal{L}_{\text{2D hyd}}^{\prime},\label{eq:new_lagrangean}
\end{equation}
 with
\begin{multline}
\mathcal{L}_{\text{2D hyd}}^{\prime}=\delta(z)\Biggl(\frac{1}{2}n_{0}m\left(\partial_{t}\bm{\upsilon}\right)^{2}\\
-\frac{1}{2}n_{0}m\beta^{2}\left(\nabla\cdot\bm{\upsilon}\right)^{2}-n_{0}e\partial_{t}\bm{\upsilon}\cdot\mathbf{A}.\Biggr)
\end{multline}
This new Lagrangean is equivalent to the Eq.~\eqref{eq:lagrangean_EM+hydro}
as both lead to the same dynamics. Applying the Euler-Lagrange equations
to Eq.~\eqref{eq:new_lagrangean} with respect to $\mathbf{A}$ leads
to Eq.~\eqref{eq:Ampere}, while the equation obtained for $\bm{\upsilon}$
reads
\begin{equation}
-m\partial_{t}^{2}\bm{\upsilon}+m\beta^{2}\nabla\left(\nabla\cdot\bm{\upsilon}\right)=e\mathbf{E}(z=0),\label{eq:fluid_eom}
\end{equation}
which can be obtained by combining Eqs.~\eqref{eq:continuity_2D}
and \eqref{eq:second_law}.

From the Lagrangean density Eq.~\eqref{eq:new_lagrangean}, we define
the canonical momenta conjugate to $\mathbf{A}$ and $\bm{\upsilon}$,
respectively, as
\begin{align}
\bm{\Pi} & =\frac{\partial\mathcal{L}^{\prime}}{\partial\left(\partial_{t}\mathbf{A}\right)}=\epsilon_{0}\bar{\epsilon}_{d}\partial_{t}\mathbf{A},\label{eq:conjugate_A}\\
\bm{\pi} & =\frac{\partial\mathcal{L}^{\prime}}{\partial\left(\partial_{t}\bm{\upsilon}\right)}=n_{0}m\partial_{t}\bm{\upsilon}-n_{0}e\mathbf{A}(z=0).\label{eq:conjugate_v}
\end{align}
In terms of the variables $\mathbf{A}$, $\bm{\Pi}$, $\bm{\upsilon}$
and $\bm{\pi}$, the classical Hamiltonian obtained from Eq.~\eqref{eq:new_lagrangean}
is given by
\begin{multline}
H=\int d^{3}\mathbf{r}\left(\frac{1}{2\epsilon_{0}}\bm{\Pi}\cdot\bar{\epsilon}_{d}^{-1}\cdot\bm{\Pi}+\frac{1}{2\mu_{0}}\left(\nabla\times\mathbf{A}\right)^{2}\right)\\
+\int d^{2}\mathbf{x}\left(\frac{\left(\bm{\pi}+en_{0}\mathbf{A}(z=0)\right)^{2}}{2n_{0}m}+\frac{1}{2}n_{0}m\beta^{2}\left(\nabla\cdot\bm{\upsilon}\right)^{2}\right).\label{eq:Hamiltonian_hydro_classical}
\end{multline}

\subsection{Canonical quantization of hydrodynamic model}

In order to quantize the classical Hamiltonian Eq.~\eqref{eq:Hamiltonian_hydro_classical},
we start by introducing the eigenmodes of the coupled electronic fluid
+ electromagnetic field. Assuming, we have in-plane translational
invariance, we write the vector potential and fluid displacement variables
as
\begin{align}
\mathbf{A}(\mathbf{r},t) & =\frac{1}{\sqrt{S}}\sum_{\mathbf{q},\lambda}\alpha_{\mathbf{q},\lambda}(t)e^{i\mathbf{q}\cdot\mathbf{x}}\mathbf{A}_{\mathbf{q},\lambda}(z)+\text{c.c.},\label{eq:mode_expansion_A}\\
\bm{\upsilon}(\mathbf{x},t) & =\frac{1}{\sqrt{S}}\sum_{\mathbf{q},\lambda}\alpha_{\mathbf{q},\lambda}(t)e^{i\mathbf{q}\cdot\mathbf{x}}\bm{\upsilon}_{\mathbf{q},\lambda}+\text{c.c.},\label{eq:mode_expansion_v}
\end{align}
where $S$ is the area of the graphene layer, $\alpha_{\mathbf{q},\lambda}(t)=\alpha_{\mathbf{q},\lambda}e^{-i\omega_{\mathbf{q},\lambda}t}$
are mode amplitudes with, $\omega_{\mathbf{q},\lambda}$the mode frequency,
and $\mathbf{A}_{\mathbf{q},\lambda}(z)$ and $\bm{\upsilon}_{\mathbf{q},\lambda}$
are mode-functions, which, following from Eqs.~\eqref{eq:Ampere}
and \eqref{eq:fluid_eom}, obey the equations

\begin{multline}
\left[\omega_{\mathbf{q},\lambda}^{2}\epsilon_{0}\bar{\epsilon}_{d}(z)-\frac{1}{\mu_{0}}D_{\mathbf{q}}\times D_{\mathbf{q}}\times\right]\mathbf{A}_{\mathbf{q},\lambda}(z)=\\
=-i\omega_{\mathbf{q},\lambda}\delta(z)en_{0}\bm{\upsilon}_{\mathbf{q},\lambda},\label{eq:hydro_eom_EM}
\end{multline}
\begin{equation}
mn_{0}\left[\omega_{\mathbf{q},\lambda}^{2}-\beta^{2}\mathbf{q}\otimes\mathbf{q}\right]\bm{\upsilon}_{\mathbf{q},\lambda}=i\omega_{\mathbf{q},\lambda}en_{0}\mathbf{A}_{\mathbf{q},\lambda}(0).\label{eq:hydro_eom_Fluid}
\end{equation}
where we defined the differential operator $D_{\mathbf{q}}=i\mathbf{q}+\hat{\mathbf{z}}\partial_{z}$.

From Eq.~\eqref{eq:hydro_eom_Fluid}, we can write
\begin{equation}
\bm{\upsilon}_{\mathbf{q},\lambda}=\frac{1}{en_{0}}\bar{\sigma}_{g}^{\text{hyd}}\left(\mathbf{q},\omega_{\mathbf{q},\lambda}\right)\cdot\mathbf{A}_{\mathbf{q},\lambda}(0),\label{eq:integrate_out_displacement}
\end{equation}
where $\bar{\sigma}_{g}^{\text{hyd}}(\mathbf{q},\omega)$ is the conductivity
within the hydrodynamic model, which we separate into transverse and
longitudinal components as
\begin{align}
\bar{\sigma}_{g}^{\text{hyd}}(\mathbf{q},\omega) & =\sigma_{g,T}^{\text{hyd}}(\mathbf{q},\omega)\left(\bar{1}-\frac{\mathbf{q}\otimes\mathbf{q}}{q^{2}}\right)\nonumber \\
 & +\sigma_{g,L}^{\text{hyd}}(\mathbf{q},\omega)\frac{\mathbf{q}\otimes\mathbf{q}}{q^{2}},\label{eq:conductivity_general}
\end{align}
respectively given by
\begin{align}
\sigma_{g,T}^{\text{hyd}}(\mathbf{q},\omega) & =\frac{e^{2}n_{0}}{m}\frac{i}{\omega},\label{eq:hydro_conductivity_transverse}\\
\sigma_{g,L}^{\text{hyd}}(\mathbf{q},\omega) & =\frac{e^{2}n_{0}}{m}\frac{i\omega}{\omega^{2}-\beta^{2}q^{2}},\label{eq:hydro_conductivity_longitudinal}
\end{align}
where we identify $D=e^{2}n_{0}/m$ as the Drude weight, which for
graphene is given by $D=e^{2}v_{F}k_{F}/\pi$. Notice that in the
the limit $\mathbf{q}\rightarrow0$, $\bar{\sigma}_{g}^{\text{hyd}}(\mathbf{q},\omega)$
reduces to the Drude model. Inserting Eq.~\eqref{eq:integrate_out_displacement}
into Eq.~\eqref{eq:hydro_eom_EM}, we obtain
\begin{equation}
D_{\mathbf{q}}\times D_{\mathbf{q}}\times\mathbf{A}_{\mathbf{q},\lambda}(z)=\frac{\omega_{\mathbf{q},\lambda}^{2}}{c^{2}}\bar{\epsilon}_{r}\left(\mathbf{q},z,\omega_{\lambda}\right)\mathbf{A}_{\mathbf{q},\lambda}(z),
\end{equation}
with the dieletric function, including screening by graphene electrons,
being given by
\begin{equation}
\bar{\epsilon}_{r}\left(\mathbf{q},z,\omega\right)=\bar{\epsilon}_{d}(z)+\frac{i}{\epsilon_{0}\omega}\bar{\sigma}_{g}^{\text{hyd}}\left(\mathbf{q},\omega\right)\delta(z),\label{eq:dieletric_function_hydro}
\end{equation}
in agreement with Eq.~\eqref{eq:mode-equation}.

Inserting the expansions Eq.~\eqref{eq:mode_expansion_A} and \eqref{eq:mode_expansion_v}
into Eq.~\eqref{eq:Hamiltonian_hydro_classical}, and using the orthogonality
properties of the mode-functions $\left(\mathbf{A}_{\mathbf{q},\lambda}(z),\,\bm{\upsilon}_{\mathbf{q},\lambda}\right)$
it is possible to write the Hamiltonian for the hydrodynamic model
as (see Appendix~\eqref{appx:Diagonalization_hydro_Hamiltonian})
\begin{equation}
H=\frac{1}{2}\sum_{\mathbf{q}\lambda}2\omega_{\mathbf{q},\lambda}^{2}\epsilon_{0}L_{\mathbf{q},\lambda}\alpha_{\mathbf{q},\lambda}^{*}\alpha_{\mathbf{q},\lambda}+\text{c.c},\label{eq:classical_hamiltonian_mode}
\end{equation}
with the mode-length defined here as
\begin{align}
L_{\mathbf{q},\lambda} & =\int dz\mathbf{A}_{\mathbf{q},\lambda}^{*}(z)\cdot\bar{\epsilon}_{d}(z)\cdot\mathbf{A}_{\mathbf{q},\lambda}(z)+\frac{n_{0}m}{\epsilon_{0}}\bm{\upsilon}_{\mathbf{q},\lambda}^{*}\cdot\bm{\upsilon}_{\mathbf{q},\lambda}\nonumber \\
 & +\frac{ien_{0}}{2\epsilon_{0}\omega_{\mathbf{q},\lambda}}\left(\mathbf{A}_{\mathbf{q},\lambda}^{*}(0)\cdot\bm{\upsilon}_{\mathbf{q},\lambda}-\bm{\upsilon}_{\mathbf{q},\lambda}^{*}\cdot\mathbf{A}_{\mathbf{q},\lambda}(0)\right).\label{eq:hydro_mode_length}
\end{align}
Promoting the mode amplitudes to quantum mechanical operators as
\begin{align}
\alpha_{\mathbf{q},\lambda}^{*}(t) & \rightarrow\sqrt{\frac{\hbar}{2\epsilon_{0}\omega_{\mathbf{q},\lambda}L_{\mathbf{q},\lambda}}}\hat{a}_{\mathbf{q},\lambda}^{\dagger}(t),\\
\alpha_{\mathbf{q},\lambda}(t) & \rightarrow\sqrt{\frac{\hbar}{2\epsilon_{0}\omega_{\mathbf{q},\lambda}L_{\mathbf{q},\lambda}}}\hat{a}_{\mathbf{q},\lambda}(t),
\end{align}
where $\hat{a}_{\mathbf{q},\lambda}^{\dagger}$$\left(\hat{a}_{\mathbf{q},\lambda}\right)$
are creation (annihilation) operators for the plasmon-polaritons,
satisfying the usual equal-time commutation relations Eq.~\eqref{eq:commutator_a_a_dagger},
the quantum Hamiltonian for the hydrodynamic model becomes
\begin{equation}
\hat{H}=\frac{1}{2}\sum_{\mathbf{q}\lambda}\hbar\omega_{\mathbf{q},\lambda}\left[\hat{a}_{\mathbf{q},\lambda}^{\dagger}\hat{a}_{\mathbf{q},\lambda}+\hat{a}_{\mathbf{q},\lambda}\hat{a}_{\mathbf{q},\lambda}^{\dagger}\right].
\end{equation}

We will now see that Eq.~\eqref{eq:hydro_mode_length} can be recast
in the same form as Eq.~\eqref{eq:hydro_mode_length}. In order to
do so, we use Eq.~\eqref{eq:integrate_out_displacement} which allows
to write Eq.~\eqref{eq:hydro_mode_length} as
\begin{multline}
L_{\mathbf{q},\lambda}=\int dz\mathbf{A}_{\mathbf{q},\lambda}^{*}(z)\cdot\bar{\epsilon}_{d}(z)\cdot\mathbf{A}_{\mathbf{q},\lambda}(z)\\
+\frac{e^{2}}{\epsilon_{0}n_{0}}\frac{\beta q^{2}}{\left(\omega^{2}-\beta q^{2}\right)^{2}}\frac{e^{2}}{\epsilon_{0}n_{0}}\mathbf{A}_{\mathbf{q},\lambda}^{*}(0)\cdot\frac{\mathbf{q}\otimes\mathbf{q}}{q^{2}}\cdot\mathbf{A}_{\mathbf{q},\lambda}(0).
\end{multline}
It is easy to see that the above equation can also be written as
\begin{multline}
L_{\mathbf{q},\lambda}=\int dz\mathbf{A}_{\mathbf{q}\lambda}^{*}(z)\cdot\Biggl(\bar{\epsilon}_{r}(\mathbf{q},z,\omega_{\lambda})\\
+\frac{\omega_{\lambda}}{2}\left.\frac{\partial}{\partial\omega}\bar{\epsilon}_{r}(\mathbf{q},z,\omega)\right|_{\omega=\omega_{\lambda}}\Biggr)\cdot\mathbf{A}_{\mathbf{q},\lambda}(z),\label{eq:mode-length_hydro}
\end{multline}
with $\bar{\epsilon}_{r}(\mathbf{q},z,\omega)$ given by Eq.~\eqref{eq:dieletric_function_hydro}.

\section{Dispersion relations and mode-functions of graphene plasmon in two
different structures\label{sec:plasmon_dispersion}}

We now wish to determine the dispersion relation and mode-functions
of graphene plasmons in two different structures (see Fig. \ref{fig:three_systems}):
a single graphene layer and a graphene sheet in the vicinity of a
perfect metal. As in the previous section, we make use of the in-plane
translational invariance of the structures being considered. Therefore,
the non-linear eigenvalue problem for the mode-functions, Eq.~\ref{eq:mode-equation},
can be written as
\begin{equation}
D_{\mathbf{q}}\times D_{\mathbf{q}}\times\mathbf{A}_{\mathbf{q},\lambda}(z)=\frac{\omega_{\mathbf{q},\lambda}^{2}}{c^{2}}\bar{\epsilon}_{r}\left(\mathbf{q},z,\omega_{\lambda}\right)\mathbf{A}_{\mathbf{q},\lambda}(z),\label{eq:eigenvalue_problem}
\end{equation}
where
\begin{equation}
\bar{\epsilon}_{r}\left(\mathbf{q},z,\omega_{\lambda}\right)=\bar{\epsilon}_{d}\left(z\right)+\sum_{\ell}i\frac{\bar{\sigma}_{g\ell}(\mathbf{q},\omega)}{\epsilon_{0}\omega}\delta(z-z_{\ell}),\label{eq:dielectric_function}
\end{equation}
where $\bar{\epsilon}_{d}\left(z\right)$ is the dielectric constant
of the medium, which we assume to be isotropic and a piecewise homogeneous
function of $z$, and $\ell$ labels the graphene layers which are
located at the planes $z=z_{\ell}$, with conductivity$\bar{\sigma}_{g\ell}(\mathbf{q},\omega)$.
We model the conductivity of each graphene layer with Eq.~\eqref{eq:conductivity_general},
which when including losses becomes
\begin{align}
\sigma_{g,T}(\mathbf{q},\omega) & =D\frac{i}{\omega+i\gamma},\label{eq:hydro_conductivity_transverse-1}\\
\sigma_{g,L}(\mathbf{q},\omega) & =D\frac{i\omega}{\omega^{2}+i\omega\gamma-\beta^{2}q^{2}},\label{eq:hydro_conductivity_longitudinal-1}
\end{align}
where $D=4E_{F}\sigma_{0}/\left(\pi\hbar\right)$ is the Drude weight,
with $E_{F}$ graphene's Fermi energy, $\sigma_{0}=\pi e^{2}/\left(2h\right)$
and $\gamma$ is a relaxation rate. When determining mode-functions
we will set $\gamma=0$, but we allow for $\gamma\neq0$, for the
situation when ohmic losses are included in Section~\eqref{sec:Plasmon_emission_rate}.
The presence of the $\delta$-functions in Eq.~\eqref{eq:dielectric_function},
implies that boundary conditions at each interface located at $z=z_{\ell}$:
\begin{align}
\hat{\mathbf{z}}\times\left(\mathbf{E}_{\mathbf{q},\lambda}\left(z_{\ell}^{+}\right)-\mathbf{E}_{\mathbf{q},\lambda}\left(z_{\ell}^{-}\right)\right) & =0,\label{eq:BC_E-1}\\
\hat{\mathbf{z}}\times\left(\mathbf{B}_{\mathbf{q},\lambda}\left(z_{\ell}^{+}\right)-\mathbf{B}_{\mathbf{q},\lambda}\left(z_{\ell}^{-}\right)\right) & =\mu_{0}\mathbf{J}_{s,\mathbf{q},\lambda}(z_{\ell}),\label{eq:BC_H-1}
\end{align}
where $\mathbf{E}_{\mathbf{q},\lambda}\left(z\right)=i\omega_{\mathbf{q},\ell}\mathbf{A}_{\mathbf{q},\lambda}(z)$
and $\mathbf{B}_{\mathbf{q},\lambda}\left(z\right)=D_{\mathbf{q}}\times\mathbf{A}_{\mathbf{q},\lambda}(z)$
are the electric and magnetic fields corresponding to mode $\mathbf{A}_{\mathbf{q},\lambda}(z)$,
and $\mathbf{J}_{s,\mathbf{q},\lambda}(z_{\ell})=\bar{\sigma}_{g\ell}(\mathbf{q},\omega)\cdot\mathbf{E}_{\mathbf{q},\lambda}\left(z_{\ell}\right)$
is the surface current in the graphene layer $\ell$. In addition
to the boundary conditions Eqs.~\eqref{eq:BC_E-1} and \eqref{eq:BC_H-1},
to determine of the plasmon modes one must impose that the fields
decay for $z\rightarrow\pm\infty$. Having determined the plasmon
mode-function, $\mathbf{A}_{\mathbf{q},\text{sp}}(z)$, and dispersion,
$\omega_{\mathbf{q},\text{sp}}$, the mode-length can be obtained
from Eqs.~\eqref{eq:mode-length_hydro} and \eqref{eq:dielectric_function}
as
\begin{multline}
L_{\mathbf{q},\text{sp}}=\int dz\epsilon_{d}(z)\mathbf{A}_{\mathbf{q},\text{sp}}^{*}(z)\cdot\mathbf{A}_{\mathbf{q},\text{sp}}(z)\\
+\frac{i}{\epsilon_{0}\omega_{\mathbf{q},\lambda}}\sum_{\ell}\mathbf{A}_{\mathbf{q},\text{sp}}^{*}(z_{\ell})\cdot\frac{\partial}{\partial\omega}\left[\omega\bar{\sigma}_{g\ell}(\mathbf{q},\omega)\right]_{\omega=\omega_{\mathbf{q},\text{sp}}}\cdot\mathbf{A}_{\mathbf{q},\text{sp}}(z_{\ell}).\label{eq:mode-length_plasmon}
\end{multline}
We will now analyse the different structures in detail.

\subsection{Single layer graphene}

We first discuss the simplest case of a single graphene sheet (see
left panel of Fig. \ref{fig:three_systems}). The problem of finding
the spectrum of the surface plasmons in a graphene sheet was first
considered in \citep{gr-spp-Jablan2009-prb} and a detailed derivation
can be found in Refs. \citep{Bludov2013,goncalves2016}. We assume
that the single layer of graphene is located at $z=0$, with a encapsulating
dielectric medium $n=2$ for $z>0$ and a medium $n=1$ for $z<0$,
such that
\begin{equation}
\epsilon_{d}(z)=\begin{cases}
\epsilon_{2}, & z>0\\
\epsilon_{1}, & z<0
\end{cases}.
\end{equation}
In order to determine the plasmon mode, we look for $p$-polarized
solutions of the electric field (the electric field lies in the plane
of incidence) in the form of evanescent waves for $z\rightarrow\pm\infty$.
In each piecewise homogeneous region we have that $D_{\mathbf{q}}\cdot\mathbf{E}_{\mathbf{q},\lambda}(z)=0$.
The mode-function must then take the form
\begin{equation}
\mathbf{A}_{\mathbf{q},\text{sp}}(z)=\begin{cases}
A_{2}^{+}\mathbf{u}_{2,\mathbf{q}}^{+}e^{-\kappa_{2,\mathbf{q}}z}, & z>0\\
A_{1}^{-}\mathbf{u}_{1,\mathbf{q}}^{-}e^{\kappa_{1,\mathbf{q}}z}, & z<0
\end{cases},\label{eq:electric_field_slg}
\end{equation}
where 
\begin{equation}
\kappa_{n,\mathbf{q}}=\sqrt{q^{2}-\frac{\omega_{\mathbf{q},\text{sp}}^{2}}{c_{n}^{2}}},
\end{equation}
with $c_{n}=c/\sqrt{\epsilon_{n}}$ the speed of light in medium $n$,
and we introduced the vectors
\begin{equation}
\mathbf{u}_{n,\mathbf{q}}^{\pm}=i\frac{\mathbf{q}}{q}\mp\frac{q}{\kappa_{n,\mathbf{q}}}\hat{\mathbf{z}}.
\end{equation}
Imposing the boundary conditions Eqs.~\eqref{eq:BC_E-1} and \eqref{eq:BC_H-1},
we obtain the following implicit relation for the surface plasmon
dispersion: 
\begin{equation}
\frac{\epsilon_{1}}{\kappa_{1,\mathbf{q}}}+\frac{\epsilon_{2}}{\kappa_{2,\mathbf{q}}}+i\frac{\sigma_{g}(\mathbf{q},\omega_{\mathbf{q},\text{sp}})}{\epsilon_{0}\omega_{\mathbf{q},\text{sp}}}=0,\label{eq:dispersion_1sheet}
\end{equation}
In general, Eq.~\eqref{eq:dispersion_1sheet} has no analytical solution,
except in the simple case where $\epsilon_{1}=\epsilon_{2}$, in which
case its solution reduces to solving a quadratic equation.

The corresponding mode-function can be written as
\begin{equation}
\mathbf{A}_{\mathbf{q},\text{sp}}(z)=\begin{cases}
\mathbf{u}_{2,\mathbf{q}}^{+}e^{-\kappa_{2,\mathbf{q}}z}, & z>0\\
\mathbf{u}_{1,\mathbf{q}}^{-}e^{\kappa_{1,\mathbf{q}}z}, & z<0
\end{cases},
\end{equation}
and the mode-length is obtained according to Eq.~\eqref{eq:mode-length_plasmon}
as
\begin{align}
L_{\mathbf{q},\text{sp}} & =\frac{\epsilon_{2}}{2\kappa_{2}^{3}}\left(\kappa_{2}^{2}+q^{2}\right)+\frac{\epsilon_{1}}{2\kappa_{1}^{3}}\left(\kappa_{1}^{2}+q^{2}\right)\nonumber \\
 & +\frac{D}{\epsilon_{0}}\frac{\beta^{2}q^{2}}{\left(\omega^{2}-\beta^{2}q^{2}\right)^{2}},\label{eq:mode-length_SingleLayer}
\end{align}
where the last term is due to the dispersion in the graphene layer.
We point out, that within the hydrodynamic model used for conductivity
of graphene, this contribution is only non-zero if non-local effects
are also included, that is if $\beta\neq0$.

\subsection{Graphene-metal structure\label{subsec:dispersion:graphene_metal}}

We now move to the more complex case of a graphene sheet near a perfect
metal (see right panel of Fig. \ref{fig:three_systems}). We assume
that the metal interface is located at $z=-d$ and the graphene layer
is located at $z=0$. The dieletric constant is given by
\begin{equation}
\epsilon_{d}(z)=\begin{cases}
\epsilon_{2}, & z>0\\
\epsilon_{1}, & 0>z>-d
\end{cases}.
\end{equation}
Noticing that the plasmon field should decay for $z\rightarrow\infty$,
the mode-function should have the form
\begin{equation}
\mathbf{A}_{\mathbf{q},\text{sp}}(z)=\begin{cases}
A_{2}^{+}\mathbf{u}_{2,\mathbf{q}}^{+}e^{-\kappa_{2}z} & z>0\\
A_{1}^{+}\mathbf{u}_{1,\mathbf{q}}^{+}e^{-\kappa_{1}z}+A_{1}^{-}\mathbf{u}_{1,\mathbf{q}}^{-}e^{\kappa_{1}z} & 0>z>-d
\end{cases}.
\end{equation}
Notice that the presence of a perfect metal at $z=-d$ implies that
the tangential component of the electric field should vanish. Imposing
the previous condition and the boundary conditions Eqs.~\eqref{eq:BC_E-1}
and \eqref{eq:BC_H-1} at $z=0$, we obtain a homogeneous system of
equations
\begin{equation}
\left[\begin{array}{ccc}
1 & -1 & -1\\
\xi_{2} & -\frac{\epsilon_{1}}{\kappa_{1}} & \frac{\epsilon_{1}}{\kappa_{1}}\\
0 & e^{\kappa_{1}d} & e^{-\kappa_{1}d}
\end{array}\right]\left[\begin{array}{c}
A_{2}^{+}\\
A_{1}^{+}\\
A_{1}^{-}
\end{array}\right]=0,\label{Eq:dispersion_metal}
\end{equation}
where $\xi_{2}=\frac{\epsilon_{2}}{\kappa_{2}}+i\frac{\sigma_{g,L}}{\epsilon_{0}\omega}$.
The dispersion relation of the screened plasmons is obtained by looking
for the zero the determinant of the previous matrix, which leads to
the condition for the dispersion relation
\begin{equation}
\frac{\epsilon_{1}}{\kappa_{1}}\coth\left(\kappa_{1}d\right)+\frac{\epsilon_{2}}{\kappa_{2}}+i\frac{\sigma_{g}}{\omega\epsilon_{0}}=0,\label{Eq:dispersion_metal_2}
\end{equation}
The boundary conditions imply that the mode-function is given by
\[
\mathbf{A}_{\mathbf{q},\text{sp}}(z)=\begin{cases}
\sinh\left(\kappa_{1}d\right)\mathbf{u}_{2,\mathbf{q}}^{+}e^{-\kappa_{2}z} & ,z>0\\
i\frac{\mathbf{q}}{q}\sinh\left(\kappa_{1}\left(z+d\right)\right)+\\
\hfill+\frac{q}{\kappa_{1,\mathbf{q}}}\hat{\mathbf{z}}\cosh\left(\kappa_{1}\left(z+d\right)\right) & ,0>z>-d
\end{cases}
\]
The mode-length can be determined from Eq.~\ref{eq:mode-length_plasmon},
and we obtain
\begin{multline}
L_{\text{sp},\mathbf{q}}=\frac{\epsilon_{2}}{2\kappa_{2,\mathbf{q}}^{3}}\sinh^{2}\left(\kappa_{1,\mathbf{q}}d\right)\left(\kappa_{2}^{2}+q^{2}\right)\\
+\frac{\epsilon_{1}}{2\kappa_{1,\mathbf{q}}^{3}}\left[\frac{1}{2}\sinh\left(2\kappa_{1,\mathbf{q}}d\right)\left(\kappa_{1,\mathbf{q}}^{2}+q^{2}\right)+d\kappa_{1,\mathbf{q}}\epsilon_{1}\frac{\omega^{2}}{c^{2}}\right]\\
+\sinh^{2}\left(\kappa_{1,\mathbf{q}}d\right)\frac{D}{\epsilon_{0}}\frac{\beta^{2}q^{2}}{\left(\omega^{2}-\beta^{2}q^{2}\right)^{2}},\label{eq:mode-length_GrapheneMetal}
\end{multline}
where, as in Eq.~\eqref{eq:mode-length_SingleLayer}, the last term
is due to the dispersion of graphene.

\begin{figure}
\centering{}\includegraphics[width=8cm]{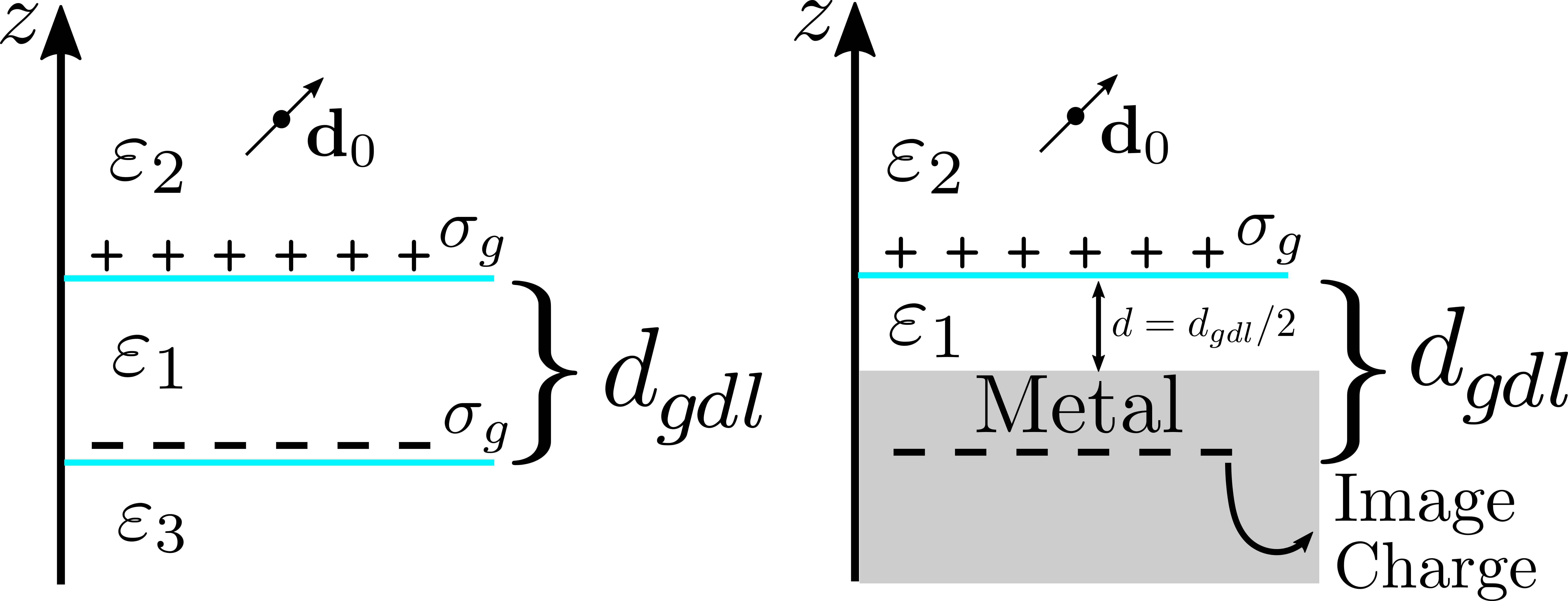}\caption{\label{fig:mirror}Comparison of the double layer graphene system
and the graphene-metal case. Due to the image charges in the metal-graphene
structure, the spectrum of the double layer graphene is equivalent
to that of the graphene-metal system if we take into account that
graphene is at at distance $d_{\text{gdl}}/2$ from the metal, where
$d_{\text{gdl}}$ is the interlayer distance in the double layer case.
Therefore, the graphene-metal distance has to be $d/2$ to obtain
the same spectrum as for the graphene double layer . For a full numerical
equivalence of the two spectra it is also necessary to have $\epsilon_{1}=\epsilon_{3}$
in the double layer geometry.}
\end{figure}

We note that the dispersion relation for the screened plasmons, Eq.~\eqref{Eq:dispersion_metal_2},
is the same one that would be obtained for the acoustic plasmons in
a symmetric graphene double layer structure (center panel of Fig.~\ref{fig:three_systems}).
In this structure, we have two graphene layers located at $z=0$ and
$z=-d_{\text{gdl}}$. The dieletric constant of the encapsulating
medium is given by
\begin{equation}
\epsilon_{d}(z)=\begin{cases}
\epsilon_{2}, & z>0\\
\epsilon_{1}, & 0>z>-d_{\text{gdl}}\\
\epsilon_{3}, & -d_{\text{gdl}}>z
\end{cases},
\end{equation}
Since the plasmonic modes should decay for $z\rightarrow\pm\infty$,
the plasmon mode-function must have the form
\begin{equation}
\mathbf{A}_{\mathbf{q},\text{sp}}(z)=\begin{cases}
A_{2}^{+}\mathbf{u}_{2,\mathbf{q}}^{+}e^{-\kappa_{2}z} & z>0\\
A_{1}^{+}\mathbf{u}_{1,\mathbf{q}}e^{-\kappa_{1}z}+A_{1}^{-}\mathbf{u}_{1,\mathbf{q}}e^{\kappa_{1}z} & 0>z>-d_{\text{gdl}}\\
A_{3}^{-}\mathbf{u}_{3,\mathbf{q}}^{-}e^{\kappa_{3}z} & -d_{\text{gdl}}>z
\end{cases}.
\end{equation}
Imposing the boundary conditions Eqs.~\eqref{eq:BC_E-1} and \eqref{eq:BC_H-1}
at $z=0$ and $z=d$, we obtain the following homogeneous system of
equations
\begin{equation}
\left[\begin{array}{cccc}
1 & -1 & -1 & 0\\
\xi_{2} & -\frac{\epsilon_{1}}{\kappa_{1}} & \frac{\epsilon_{1}}{\kappa_{1}} & 0\\
0 & e^{\kappa_{1}d} & e^{-\kappa_{1}d} & -e^{-\kappa_{3}d}\\
0 & \frac{\epsilon_{1}}{\kappa_{1}}e^{\kappa_{1}d_{\text{gdl}}} & -\frac{\epsilon_{1}}{\kappa_{1}}e^{-\kappa_{1}d_{\text{gdl}}} & \xi_{3}e^{-\kappa_{3}d_{\text{gdl}}}
\end{array}\right]\left[\begin{array}{c}
A_{2}^{+}\\
A_{1}^{+}\\
A_{1}^{-}\\
A_{3}^{-}
\end{array}\right]=0,\label{eq:Double_linear_sys}
\end{equation}
where $\xi_{2}=\frac{\epsilon_{2}}{\kappa_{2}}+i\frac{\sigma_{g\text{top},L}}{\epsilon_{0}\omega}$
and $\xi_{3}=\frac{\epsilon_{3}}{\kappa_{3}}+i\frac{\sigma_{g\text{bot},L}}{\epsilon_{0}\omega}$.
The dispersion relation is obtained from zeroing the determinant of
the previous matrix. Since the system is composed of two graphene
sheets the double layer structure has two dispersion branches, a low
energy one -- the acoustic mode -- and a high energy one -- the
optical mode. In the optical mode, the charge oscillations in the
two graphene sheets are in-phase; whereas in the acoustic mode, the
charge oscillations are out-of-phase. In the particular case where,
we have a symmetry structure, with $\epsilon_{2}=\epsilon_{3}$ and
the conductivities of the top and bottom graphene layers are the same
$\sigma_{g\text{top},L}=\sigma_{g\text{bot},L}=\sigma_{g,L}$, the
zeroing of the determinant factorizes into two independent expressions
\begin{equation}
\frac{\epsilon_{1}}{\kappa_{1}}\tanh\left(\frac{\kappa_{1}d_{\text{dlg}}}{2}\right)+\frac{\epsilon_{2}}{\kappa_{2}}+i\frac{\sigma_{g,L}}{\omega\epsilon_{0}}=0\label{eq:optico}
\end{equation}
for the optical mode, and 
\begin{equation}
\frac{\epsilon_{1}}{\kappa_{1}}\coth\left(\frac{\kappa_{1}d_{\text{dlg}}}{2}\right)+\frac{\epsilon_{2}}{\kappa_{2}}+i\frac{\sigma_{g,L}}{\omega\epsilon_{0}}=0\label{eq:acustico}
\end{equation}
for the acoustic one. Notice that the relation for the acoustic mode
dispersion Eq.~\eqref{eq:acustico} coincides with the equation for
the screened plasmon Eq.~\eqref{Eq:dispersion_metal_2} provided
$d_{\text{dlg}}=2d$. This fact can be understood in terms of image
charges as depicted in Fig.~\ref{fig:mirror}.

\begin{figure}
\begin{centering}
\includegraphics[clip,width=8cm]{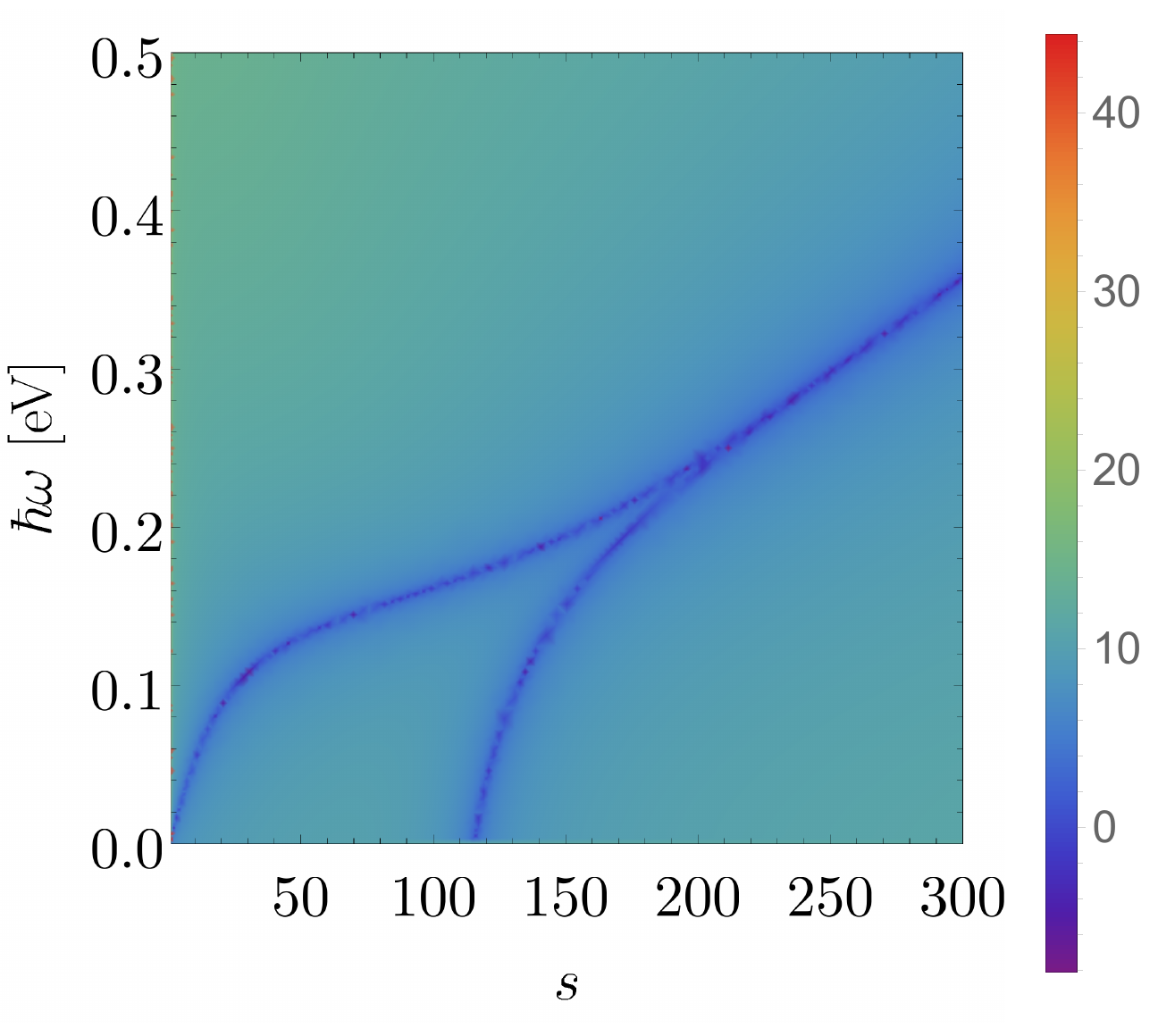}
\par\end{centering}
\centering{}\includegraphics[clip,width=8cm]{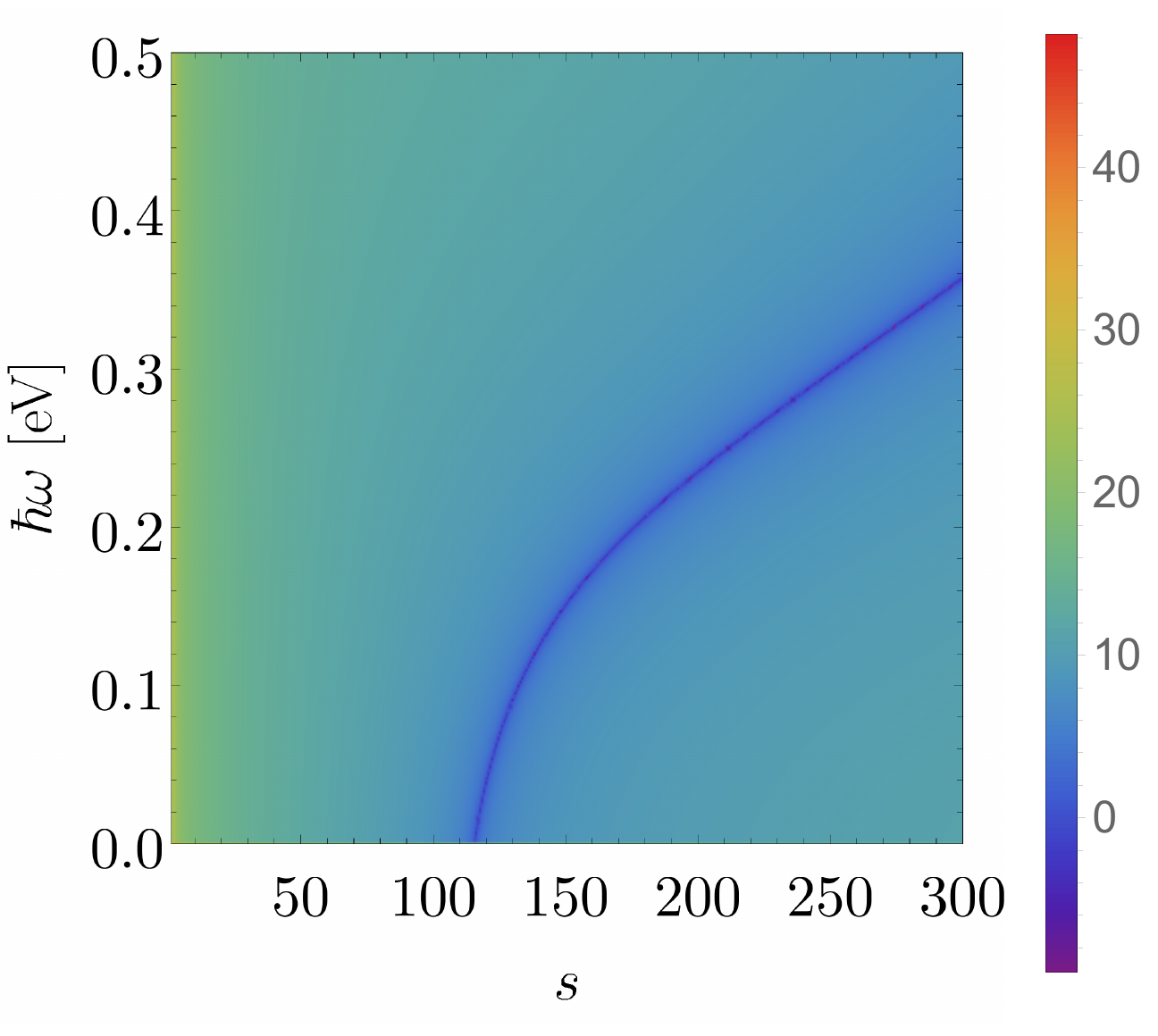}\caption{\label{fig:two_dispersions}Loss function {[}Eq.~\eqref{eq:loss}{]}
for the $p$-polarization reflection coefficient for: (Top panel)
a graphene double layer structure; (Bottom panel) graphene near a
perfect metal. The parameters use in the top panel are $d_{\text{gdl}}=20$
nm, $\epsilon_{3}=1$, and for the bottom panel $d=10$ nm. The remaining
parameters used in both panels are: $\epsilon_{1}=3.9$, $\epsilon_{2}=1$,
$E_{F}=0.2$ eV and $\hbar\gamma=10$ meV. In both plots, non-local
effects were neglected ($\beta=0$).}
\end{figure}

In the bottom panel of Fig. \ref{fig:two_dispersions} we depict the
loss function for the graphene-metal system. Clearly, only one branch
is seen, which coincides with the acoustic branch of the double layer
graphene upon considering $d$ equal to half that of the double layer
system, as explained in Fig.~\ref{fig:mirror}.

An alternative way to obtain the plasmon dispersion relation is to
look for poles (or resonances in the presence of losses) in the so
called loss function\citep{goncalves2016} , which is defined as
\begin{equation}
{\cal L}(\mathbf{q},\omega)=-\text{Im}\left[r_{p}(\mathbf{q},\omega)\right],\label{eq:loss}
\end{equation}
where $r_{p}(\mathbf{q},\omega)$ is the reflection coefficient of
the structure in consideration for the $p$-polarization, and $\mathbf{q}$
and $\omega$ are the in-plane wavevector and frequency of the impingent
radiation, respectively (see Appendix \ref{Appx:reflection_coefficients}).
For a symmetric graphene double layer ($\epsilon_{1}=\epsilon_{3}$
and $\sigma_{g\text{top},L}=\sigma_{g\text{bot},L}$) and neglecting
losses $\gamma\rightarrow0$, this coefficient has poles at the solutions
of Eqs.~\eqref{eq:acustico} and \eqref{eq:optico}, as can be seen
comparing those equations with Eq.~\eqref{eq:rp_metal_graphene}.
The loss function for the double layer graphene is depicted in the
left panel of Fig.~\ref{fig:two_dispersions}, as function of $\omega$
and of a dimensionless parameter $s=qc/\omega$ which defines the
dispersion relation of the single graphene layer, clad by two different
dielectrics of dielectric functions $\epsilon_{1}$ and $\epsilon_{2}$,
in the electrostatic limit by 
\begin{equation}
\omega(s)=\frac{4\alpha E_{F}}{\epsilon_{1}+\epsilon_{2}}s,\label{eq:s}
\end{equation}
where $E_{F}$ is the Fermi energy of graphene and $\alpha$ is the
fine structure constant of vacuum. In the top panel of Fig.~\ref{fig:two_dispersions}
two branches are clearly seen: a high energy one -- the optical branch
-- and the acoustic branch at lower energies. At high energies and
high $s$ the two branches merge and converge to the single layer
branch. The reader may wonder why the lower branch starts at finite
momentum. This happens due to the definition of the $s$ parameter,
which involves both the frequency and the real wavenumber $q$. This
choice allows to clearly separate the two branches in the $(\omega,s)$
plane.

\section{Application: quantum emission close to graphene structures\label{sec:Plasmon_emission_rate}}

We will now apply the quantization of the plasmon modes in grahene
structures to the problem of spontaneous emission by a quantum emitter
which is located above the structure. We model the quantum emitter
as a two-level system embedded in medium $2$ at position $\mathbf{r}_{0}=\left(0,0,z_{0}\right)$.
The quantum emitter couples to the plasmonic field via dipolar coupling:$\hat{H}_{\text{sp-d}}=-\hat{\mathbf{d}}\cdot\hat{\mathbf{E}}_{\text{\text{sp}}}(\mathbf{r}_{0}),$
with
\begin{align}
\hat{\mathbf{d}} & =\mathbf{d}_{ge}\hat{c}_{g}^{\dagger}\hat{c}_{e}+\text{h.c.}\\
\hat{\mathbf{E}}_{\text{sp}}(\mathbf{r}) & =\sum_{\mathbf{q}}\left(i\sqrt{\frac{\hbar\omega_{\mathbf{q},\text{sp}}}{2S\epsilon_{0}L_{\mathbf{q},\text{sp}}}}\mathbf{A}_{\mathbf{q},\text{sp}}(z)e^{i\mathbf{q}\cdot\mathbf{x}}\hat{a}_{\mathbf{q},\text{sp}}+\text{h.c.}\right)
\end{align}
where $\hat{c}_{g/e}^{\dagger}$($\hat{c}_{g/e}$) is the creation
(annihilation) operator for the ground/excited state of the two-level
system and $\mathbf{d}_{ge}=\left\langle g\left|\hat{\mathbf{d}}\right|e\right\rangle $
is the dipole matrix element.

The transition rate of an emitter due to emission of surface plasmons
in graphene is given by Fermi's golden rule \citep{NovotnyHecht_book,Archambault2010}\textcolor{black}{:}
\begin{equation}
\Gamma_{\text{sp}}=\frac{2\pi}{\hbar}\sum_{\mathbf{q}}\left|\left\langle g;n_{\mathbf{q},\text{sp}}+1\right|\hat{\mathbf{d}}\cdot\hat{\mathbf{E}}\left|e;n_{\mathbf{q},\text{sp}}\right\rangle \right|^{2}\delta\left(\hbar\omega_{0}-\hbar\omega_{\mathbf{q},\text{sp}}\right),\label{eq:FermiGold}
\end{equation}
where $\omega_{0}$ is the transition frequency, $\left|g;n_{\mathbf{q},\text{sp}}+1\right\rangle $
represents a final state with one more surface plasmon and the emitter
in the ground state and $\left|e;n_{\mathbf{q},\text{sp}}\right\rangle $
represents an initial state with $n_{\mathbf{q},\text{sp}}$ plasmons
in graphene and the emitter in the excited state. The transition matrix
element for spontaneous plasmon emission, when there are no surface
plasmons in the initial state,is given by
\begin{equation}
\left\langle g;1\right|\hat{\mathbf{d}}\cdot\hat{\mathbf{E}}\left|e;0\right\rangle =i\sqrt{\frac{\hbar\omega_{\mathbf{q},\text{sp}}}{2S\epsilon_{0}L_{\mathbf{q},\text{sp}}}}\mathbf{d}_{ge}\cdot\mathbf{A}_{\mathbf{q},\text{sp}}^{*}(z_{0}),
\end{equation}
With this result the transition rate reads 
\begin{multline}
\Gamma_{\text{sp}}=\frac{1}{4\pi\hbar\epsilon_{0}}\int d^{2}\mathbf{q}\frac{\hbar\omega_{\mathbf{q},\text{sp}}}{L_{\mathbf{q},\text{sp}}}\left|\mathbf{d}_{ge}\cdot\mathbf{A}_{\mathbf{q},\text{sp}}^{*}(z_{0})\right|^{2}\times\\
\times\delta\left(\hbar\omega_{0}-\hbar\omega_{\mathbf{q},\text{sp}}\right),\label{eq:transition_rate}
\end{multline}
Using the mode-function we can write 
\begin{multline}
\left|\mathbf{d}_{ge}\cdot\mathbf{A}_{\mathbf{q},\text{sp}}^{*}(z_{0})\right|^{2}=\mathcal{N}_{\mathbf{q}}\left|\mathbf{d}_{ge}\right|^{2}e^{-2\kappa_{2,\mathbf{q}}z_{0}}\times\\
\times\left(\cos^{2}\phi\sin^{2}\psi+\frac{q^{2}}{\kappa_{2,\mathbf{q}}^{2}}\cos^{2}\psi\right),\label{eq:matrix_elment}
\end{multline}
$\psi$ is the angle the dipole makes with the axis perpendicular
to graphene ($z$-axis), and $\phi$ is the azimuthal angle. The prefactor
$\mathcal{N}_{\mathbf{q}}$ is define as $\mathcal{N}_{\mathbf{q}}=1$
for the single-layer graphene case and as $\mathcal{N}_{\mathbf{q}}=\sinh^{2}\left(\kappa_{1,\mathbf{q}}d\right)$
for the graphene+metal structure. Using the in-plane isotropy of the
system, the momentum integration in Eq.~\eqref{eq:transition_rate}
can be trivially performed, yielding:
\begin{multline}
\Gamma_{\text{sp}}=\mathcal{N}_{\mathbf{q}_{0}}\frac{q_{0}\omega_{\mathbf{q}_{0},\text{sp}}}{L_{\mathbf{q}_{0},\text{sp}}}\left(\frac{\partial\omega_{\mathbf{q}_{0},\text{sp}}}{\partial q}\right)^{-1}\frac{\left|\mathbf{d}_{ge}\right|^{2}}{4\hbar\epsilon_{0}}e^{-2\kappa_{2,\mathbf{q}_{0}}z_{0}}\times\\
\times\left(\sin^{2}\psi+2\frac{q^{2}}{\kappa_{2,\mathbf{q}_{0}}^{2}}\cos^{2}\psi\right),\label{eq:general_decay_formula}
\end{multline}
where $q_{0}$ is the momentum of a surface plasmon, with frequency
$\omega_{0}$, i.e., $\omega_{0}=\omega_{\mathbf{q}_{0},\text{sp}}$.
So far the expression for $\Gamma_{\text{sp}}$ is general. The differences
arise from the particular forms of the dispersion $\omega_{\mathbf{q},\text{sp}}$,
the mode length $L_{\mathbf{q},\text{sp}}$ and the prefactor $\mathcal{N}_{\mathbf{q}}$.

\subsection{Decay rate for local conductivity}

We will now study the plasmon emission rate, when non-local effects
are neglected, $\beta=0$ in Eq.~\eqref{eq:hydro_conductivity_longitudinal-1}.

\begin{figure}
\centering{}\includegraphics[width=8cm]{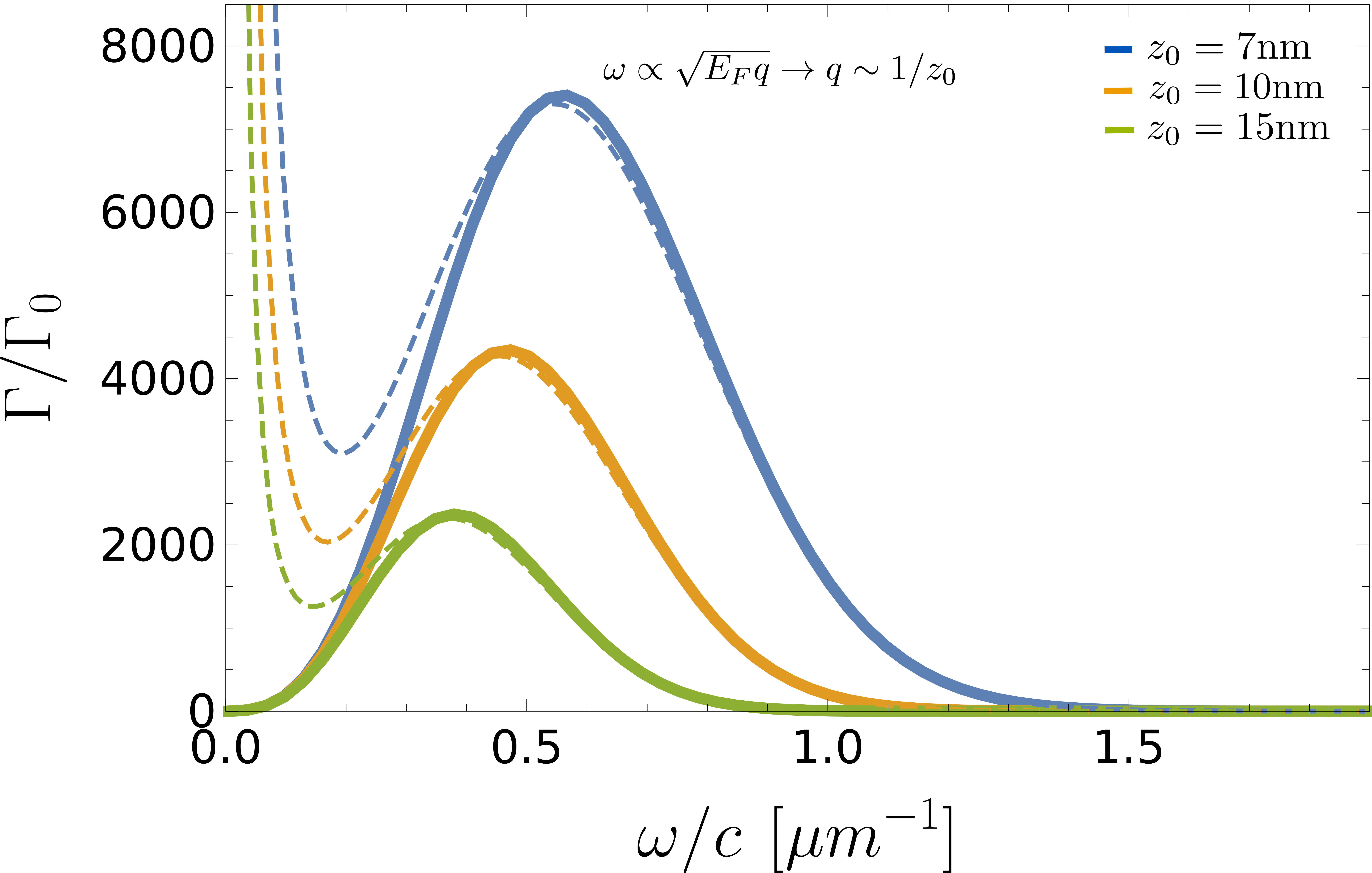}\caption{\label{fig:rate_single_layer_classical}Decay rate of a quantum emitter
close to a single graphene layer as a function of emitter frequency,
for different emitter-graphene distances $z_{0}$. The solid lines
show the decay rate due to plasmon emission as evaluated with Eq.~\eqref{eq:Gamma_final}.
The dashed lines show the total decay rate computed used Eq.~\eqref{eq:Dyadic_Green_Gamma}.
The parameters used are $\epsilon_{1}=\epsilon_{2}=1$, $E_{F}=0.3$
eV, and $\cos^{2}\psi=1/3$. For the evaluation of the total decay
rate a broadening of $\hbar\gamma=4$ meV was used. The local form
of graphene conductivity was used $(\beta=0)$.}
\end{figure}

\begin{figure}
\centering{}\includegraphics[width=8cm]{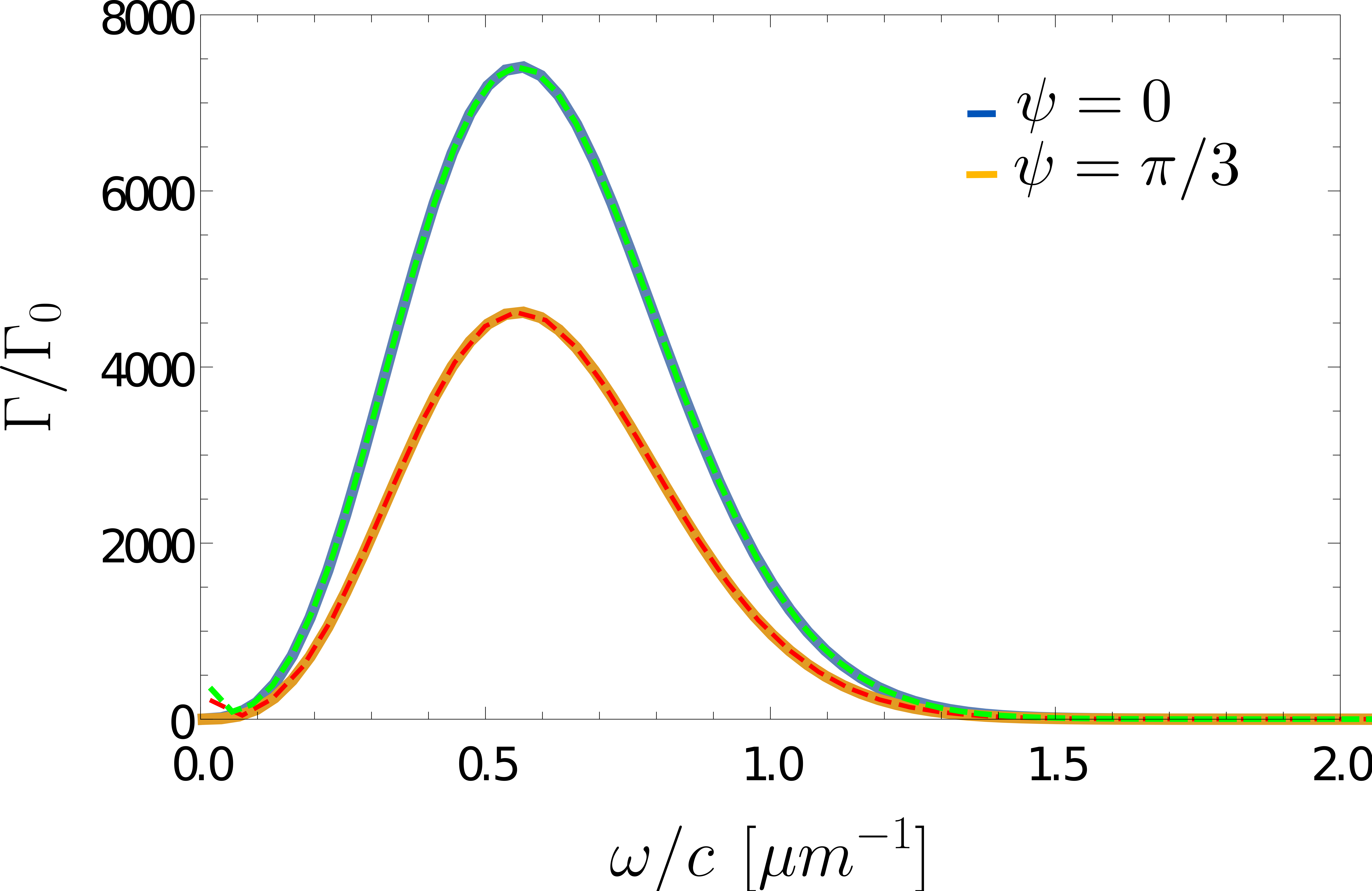} \caption{\label{fig:comparison_quantum_classical}Decay rate of a quantum emitter
close to a single graphene layer as a function of emitter frequency,
for different dipole orientations: $\psi=0$ and $\psi=\pi/3$. Solid
lines show the emission rate {[}Eq.~\eqref{eq:Gamma_final}{]} and
the dashed lines show the total decay rate {[}Eq.~\eqref{eq:Dyadic_Green_Gamma}{]}.
The parameters considered where $z_{0}=70$ nm, $E_{F}=0.4$ eV, and
$\epsilon_{1}=\epsilon_{2}=1$. In the evaluation of the total decay
rate a broadening factor of $\hbar\gamma=0.1$ meV was used. For such
small losses, the decay rate is completely dominated by the plasmon
emission. The local form of graphene conductivity was used $(\beta=0)$.}
\end{figure}

We first focus on the case of a quantum emitter close to a single
graphene layer and the same dielectric above and bellow the graphene
layer, $\epsilon_{1}=\epsilon_{2}=\epsilon$. Using the analytic solution
of Eq.~\eqref{eq:dispersion_1sheet} for this case, we can write
Eq.~\eqref{eq:general_decay_formula} as

\begin{multline}
\Gamma_{\text{sp}}^{\text{gsl}}=\frac{d_{ge}^{2}}{4\hbar\epsilon_{0}}\left[2\omega_{0}^{4}\left(\frac{\hbar}{2\alpha E_{F}c}\right)^{2}+\frac{\omega_{0}^{2}\epsilon}{c^{2}}\right]\frac{e^{-2\kappa_{\mathbf{q}}z_{0}}}{L_{\mathbf{q}_{0},\text{sp}}}\times\\
\times\left[\sin^{2}\psi+2\kappa^{-2}\left[\omega_{0}^{4}\left(\frac{\hbar}{2\alpha E_{F}c}\right)^{2}+\frac{\omega_{0}^{2}\epsilon}{c^{2}}\right]\cos^{2}\psi\right].\label{eq:Gamma_final}
\end{multline}
In Figs.~\ref{fig:rate_single_layer_classical} and\ref{fig:comparison_quantum_classical},
we plot the ratio $\Gamma_{\text{sp}}^{\text{gsl}}/\Gamma_{0}$, where
$\Gamma_{0}=d_{ge}^{2}\omega_{0}^{3}/\left(3\pi\epsilon_{0}\hbar c^{3}\right)$
the total decay rate of an emitter in vacuum, for different graphene-quantum
emitter distances $z_{0}$ and dipole orientations. For comparison,
we will display the ratio $\Gamma_{\text{full}}/\Gamma_{0}$, where
$\Gamma_{\text{full}}$ is the total decay rate (including both plasmon
and photonic losses) of the quantum emitter. The total decay rate
can be computed from the knowledge of the reflection coefficients,
which are incorporated into the dyadic EM Green's function \citep{Agarwal1975_IV,NovotnyHecht_book,goncalves2016}\textcolor{black}{{}
(see Ref. \citep{Koppens2011} for a detailed study of the properties
of an emitter near graphene using Dyadic Green's functions). For a
dipole in medium $n=2$, the total decay rate is given by }
\begin{align}
\frac{\Gamma_{\text{full}}}{\Gamma_{0}} & =1+\frac{3}{2}\cos^{2}\psi\text{Re}\left(\int_{0}^{\infty}\frac{dq}{k_{1}^{3}}\frac{q^{3}}{k_{z,1}}r_{p}e^{-2\kappa_{2}z_{0}}\right)\nonumber \\
 & +\frac{3}{4}\sin^{2}\psi\text{Re}\left(\int_{0}^{\infty}\frac{dq}{k_{1}}\frac{q}{k_{z,1}}\left(r_{s}-r_{p}\frac{c^{2}q^{2}}{\omega^{2}}\right)e^{-2\kappa_{2}z_{0}}\right),\label{eq:Dyadic_Green_Gamma}
\end{align}
where $r_{s/p}$ are reflection coefficients for the $s/p$-polarization
and $k_{n}=\omega/c_{n}$ and $k_{z,n}$ are defined in Appendix \ref{Appx:reflection_coefficients}.

A number of details are worth mentioning. There are two distinct behaviors
of the decay rate $\Gamma_{\text{full}}$. At low frequency the curves
shoot up due to Ohmic losses, which are not included in Eq.~\eqref{eq:Gamma_final}.
At intermediate frequencies, the curves develop a clear resonance
due to the excitation of surface plasmons in graphene. Also the maximum
of the resonances blue-shifts with the decrease of the distance of
the dipole to the graphene sheet. This behavior is easily explained
remembering that the dispersion of the surface plasmons in single
layer graphene is proportional to $\sqrt{q}$. Since the distance
$z_{0}$ introduces a momentum scale $q\sim1/z_{0}$, smaller $z_{0}$
values correspond to higher $q$-values and higher energies of the
resonant maximum. Equation (\ref{eq:Gamma_final}) for the plasmon
emission rate produces exactly the same resonance (same magnitude
and same maximum of the resonance position) as $\Gamma_{\text{full}}$,
indicating that in this region, the decay rate of the quantum emitter
is dominated by plasmon emission. This is further shown in Fig. \ref{fig:comparison_quantum_classical},
where losses were arbitrarily reduced in the evaluation of $\Gamma_{\text{full}}$.
We can also avoid the superposition of the Ohmic and surface plasmon
contributions choosing either a larger Fermi energy or a larger distance
from graphene to the metal. In Fig. \ref{fig:comparison_quantum_classical}
the ratio $\Gamma/\Gamma_{0}$ is smaller than the ones in Fig.~\ref{fig:rate_single_layer_classical}
due to the larger distance of the dipole to the graphene sheet.

An analytic expression for the plasmon emission rate can be also obtained
for the case of graphene-metal structure assuming that $\kappa_{1}d\ll1$.
In this limit, Eq.~\ref{Eq:dispersion_metal_2} can be approximately
solved, yielding $\hbar\omega_{\mathbf{q},\text{sp}}=\sqrt{4\alpha dE_{F}\hbar c/\epsilon}q$.
Plugging this result in Eq.~\ref{eq:general_decay_formula}, we obtain

\begin{figure}
\centering{}\includegraphics[width=8cm]{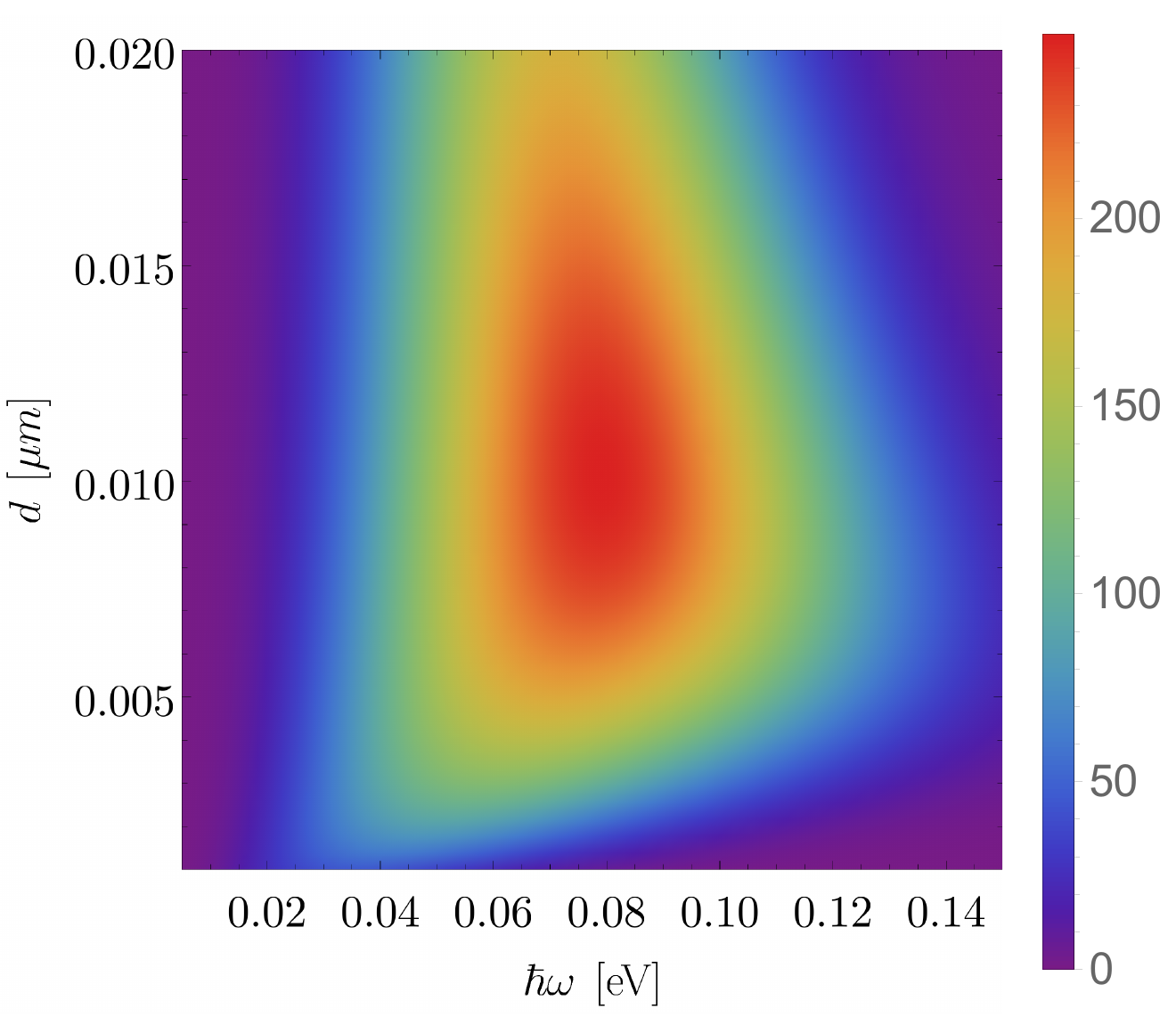}\caption{\label{fig:metal_density_plot}Density plot of the transition rate
of a quantum emitter due to the emission of surface plasmons in a
graphene-metal structure. The used parameters are $z_{0}=30$ nm,
$E_{F}=0.2$ eV, $\epsilon_{1}=3.9$, and $\epsilon_{2}=1$, $\psi=0$.
Making line cuts at constant $d$ we can easily see the presence of
a resonance in the dipole transition rate due to the excitation of
graphene screened plasmons.}
\end{figure}

\textcolor{black}{
\begin{multline}
\Gamma_{\text{sp}}^{\text{gm}}\simeq\frac{d_{12}^{2}}{4\varepsilon_{0}\hbar}\frac{\varepsilon\sinh^{2}(\kappa_{1}d)}{4\alpha dE_{F}\hbar c}\frac{(\hbar\omega_{0})^{2}e^{-2\kappa_{2}z_{0}}}{L_{sq}(\omega_{at})}\\
\times\left(\sin^{2}\psi+2\frac{q^{2}}{\kappa_{2}^{2}}\cos^{2}\psi\right).\label{eq:gamma_sp_graphene_metal}
\end{multline}
}This plasmon emission rate is shown in Fig.~\ref{fig:metal_density_plot}
as a function of $\omega$ and the graphene-metal separation $d$.
As in the case of a single graphene layer, for each $d$ there is
a well defined peak as function of $\omega$.

\subsection{Decay rate with non-local effects}

\begin{figure}
\begin{centering}
\includegraphics[width=8cm]{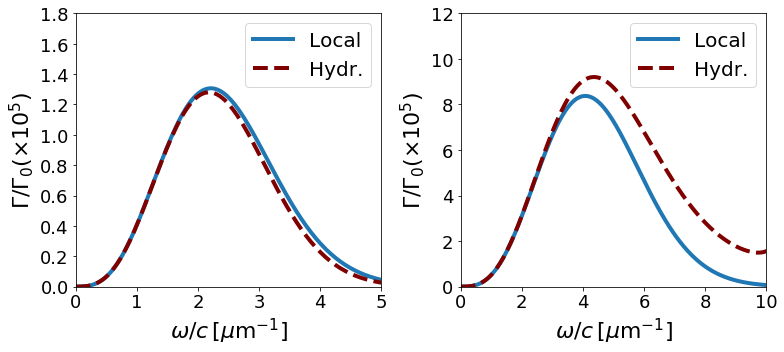}
\par\end{centering}
\caption{\label{fig:Hydro_comparison}Comparison between the plasmon emission
rate of a quantum emitter close to a single layer of graphene using
the local ($\beta=0$) and non-local (hydr.) ($\beta\protect\neq0$)
models for the graphene conductivity (see Eq.~\eqref{eq:hydro_conductivity_longitudinal}).
The used $E_{F}=0.6$ eV, $\epsilon_{1}=\epsilon_{2}=3.9$, $\psi=0$.
Left panel : $z_{0}=7$ nm. Right panel: $z_{0}=2$ nm. The nonlocal
effects causes an increase and blueshift of the resonance in the transition
rate as the distance $z_{0}$ between the quantum emitter and the
graphene decreases.}
\end{figure}

We now focus on the role of non-local effects in the graphene conductivity
play in the emitter decay rate. We will restrict ourselves to the
case of a quantum emitter close to a single layer of graphene. In
Fig.~\ref{fig:Hydro_comparison}, we compare the transition rate
of a quantum emitter calculated taking into account non-local effects
$\beta\neq0$ in Eq.~\eqref{eq:hydro_conductivity_longitudinal},
with the local case $\beta=0$. The left panel shows the result for
$E_{F}=0.6$ eV, $z_{0}=7$ nm, while the right panel shows the result
for$E_{F}=0.6$ eV, $z_{0}=2$ nm. We can see that the nonlocal effects
play an important role at smaller emitter-graphene distances. As the
distance between quantum emitter and the graphene layer, $z_{0}$,
decreases, non-locality leads to an increase and blueshift of the
resonance in the transition rate as a function of the emitter frequency.
We have also verified (not shown) that the non-local effects are also
more prominent in the case of lower electronic densities.

\section{Conclusions\label{sec:Conclusions}}

In this paper, we have performed the quantization of graphene plasmons,
in the absence of losses, and applied the field quantization to the
interaction of an emitter with doped graphene. The quantization was
performed using both a macroscopic energy approach and a quantum hydrodynamic
model, which allows for the inclusion of non-local effects in the
EM response of graphene. The quantization approaches used allow for
the determination of the plasmon EM field mode-functions and, importantly,
their normalization, which becomes non-trivial when dispersion is
included.

When comparing the total decay rate of a quantum emitter (as obtained
using the full EM dyadic Green's function) with the decay rate due
to quantum emission, it was shown that plasmon emission completely
dominates the decay rate, for typical emitter-graphene separations
and emitter frequencies. It was shown that non-local effects in the
graphene response, become increasingly important for smaller graphene-emitter
separations and smaller Fermi energies.

The advantage of the quantization method developed in this work lies
in its simplicity. The mode-functions are obtained from the solution
of the Maxwell equations for the vector potential, and the normalization
of the mode-functions can be expressed as a simply integral, which
only involves the dielectric function of the medium. For situations
when only a few modes contribute significantly to the physics, as
in the case of quantum emission dominated by plamons, mode-functions
allows to a much simpler and physically transparent description than
the full EM dyadic Green's function. Since, the determination of the
mode-functions only involves the solution of the classical Maxwell
equations, the quantization of the electromagnetic field of more complexes
structures can be easily achieved. We also note that the procedure
gives both the quantized form of the electric and magnetic fields.
Therefore, we could also study the enhancement of the spin relaxation.
The only change would be the interaction Hamiltonian, which in this
case would be of the form $H_{I}=-\bm{\mu}\cdot\mathbf{B}$, where
$\bm{\mu}$ is the magnetic dipole moment of the emitter and $\mathbf{B}$
the quantized magnetic field of the surface plasmon.

In possession of a quantized theory for graphene plasmons, we have
set the stage for the future discussion of other quantum effects involving
these collective excitation made simultaneously of light and matter.

\section*{Acknowledgments}

B.A.F. and N.M.R.P. acknowledge support from the European Commission
through the project ``Graphene-Driven Revolutions in ICT and Beyond''
(Ref. No. 785219), and the Portuguese Foundation for Science and Technology
(FCT) in the framework of the Strategic Financing UID/FIS/04650/2013.
Additionally, N. M. R. P. acknowledges COMPETE2020, PORTUGAL2020,
FEDER and the Portuguese Foundation for Science and Technology (FCT)
through projects PTDC/FIS-NAN/3668/2013 and POCI-01-0145-FEDER-028114.
B. A. acknowledges the hospitality of CeFEMA where he was a visiting
researcher during part of the time in which this work was developed.

\appendix

\section{Reflection coefficients of graphene structures\label{Appx:reflection_coefficients}}

In this appendix we provide the reflection coefficients needed for
the evaluation of the full decay rate of a quantum emitter, Eq.~\eqref{eq:Dyadic_Green_Gamma}.

\subsection{Reflection coefficients for a single graphene layer}

The Fresnel problem for a single graphene sheet has been considered
in Ref. \citep{goncalves2016}. Therefore we provide here the final
results only and for simplicity we assume we are dealing with non-magnetic
media. For an incoming wave from region $2$, and for the $s-$polarization
the reflection and transmission coefficient read

\begin{eqnarray}
r_{s} & = & \frac{k_{z,2}-k_{z,1}-\mu_{0}\omega\sigma_{g,T}}{k_{z,2}+k_{z,1}+\mu_{0}\omega\sigma_{g}},\label{eq:rs_single}\\
t_{s} & = & \frac{2k_{z,2}}{k_{z,2}+k_{z,1}+\mu_{0}\omega\sigma_{g,T}},
\end{eqnarray}
whereas for the $p-$polarization we have 
\begin{eqnarray}
r_{p} & = & -\frac{\epsilon_{2}k_{z,1}-\epsilon_{1}k_{z,2}-k_{z,1}k_{z,2}\sigma_{g,L}/\left(\omega\epsilon_{0}\right)}{\epsilon_{2}k_{z,1}+\epsilon_{1}k_{z,2}+k_{z,1}k_{z,2}\sigma_{g,L}/\left(\omega\epsilon_{0}\right)},\label{eq:rp_single}\\
t_{p} & = & \sqrt{\frac{\epsilon_{1}}{\epsilon_{2}}}\frac{2k_{z,2}\epsilon_{1}}{\epsilon_{2}k_{z,1}+\epsilon_{1}k_{z,2}+k_{z,1}k_{z,2}\sigma_{g,L}/\left(\omega\epsilon_{0}\right)}.
\end{eqnarray}
In these equations, we have $k_{z,n}=\sqrt{\epsilon_{n}\omega^{2}/c^{2}-q^{2}}=i\kappa_{n}$,
and $\sigma_{g,L/T}$ is the optical longitudinal/transverse conductivity
of graphene, $\epsilon_{0}$ and $\mu_{0}$ are the vacuum permitivity
and permeability, respectively, $q$ is the wavenumber along the graphene
sheet, and $\omega$ is the frequency of the electromagnetic radiation.

\subsection{Metal-graphene reflectance coefficient}

The reflection coefficient for the $p$-polarization for this structure
is given by 
\begin{equation}
r_{p}=1-\frac{2k_{z,1}k_{2}^{2}\sin\left(k_{z,1}d\right)}{k_{z,1}\sin\left(k_{z,1}d\right)\left(k_{z,2}\mu_{0}\omega\sigma_{g,L}+k_{2}^{2}\right)+ik_{z,2}k_{1}^{2}\cos\left(k_{z,1}d\right)},\label{eq:rp_metal_graphene}
\end{equation}
where $k_{n}^{2}=\epsilon_{n}\omega^{2}/c^{2}$, $d$ is the graphene-metal
distance, and $\kappa_{n}^{2}=k_{n}^{2}-q^{2}$, where $q$ is the
wavenumber along the graphene sheet. For the $s-$polarization we
have 
\begin{equation}
r_{s}=-1+\frac{2k_{z,2}\sin\left(k_{z,1}d\right)}{\sin\left(k_{z,1}d\right)\left(k_{z,2}+\mu_{0}\omega\sigma_{g,T}\right)+ik_{z,1}\cos\left(k_{z,1}d\right)}.
\end{equation}

\begin{widetext}

\section{Diagonalization of Hydrodynamic Hamiltonian\label{appx:Diagonalization_hydro_Hamiltonian}}

\subsection{Orthogonality of mode-functions}

We start by showing that the mode-functions $\left(\mathbf{A}_{\mathbf{q},\lambda}(z),\,\bm{\upsilon}_{\mathbf{q},\lambda}\right)$
obey certain orthogonality conditions which will be useful. We start
by writing the equations for the mode-functions within the hydrodynamic
model, Eqs\@.~\eqref{eq:hydro_eom_EM} and \eqref{eq:hydro_eom_Fluid},
in matrix form as
\begin{equation}
\left[\begin{array}{cc}
\omega_{\mathbf{q},\lambda}^{2}\epsilon_{0}\bar{\epsilon}_{d}(z)-\frac{1}{\mu_{0}}D_{\mathbf{q}}\times D_{\mathbf{q}}\times & i\omega_{\mathbf{q},\lambda}en_{0}\delta(z)\\
-i\omega_{\mathbf{q},\lambda}en_{0}\delta(z) & n_{0}m\left(\omega_{\mathbf{q},\lambda}^{2}-\beta^{2}\mathbf{q}\otimes\mathbf{q}\right)\delta(z)
\end{array}\right]\left[\begin{array}{c}
\mathbf{A}_{\mathbf{q},\lambda}(z)\\
\bm{\upsilon}_{\mathbf{q},\lambda}
\end{array}\right]=0.\label{eq:mode1}
\end{equation}
Let us now consider another mode $\lambda^{\prime}$, with mode-function
$\left(\mathbf{A}_{\mathbf{q},\lambda^{\prime}}(z),\,\bm{\upsilon}_{\mathbf{q},\lambda^{\prime}}\right)$,
solution of
\begin{equation}
\left[\begin{array}{cc}
\omega_{\mathbf{q},\lambda^{\prime}}^{2}\epsilon_{0}\bar{\epsilon}_{d}(z)-\frac{1}{\mu_{0}}D_{\mathbf{q}}\times D_{\mathbf{q}}\times & i\omega_{\mathbf{q},\lambda^{\prime}}en_{0}\delta(z)\\
-i\omega_{\mathbf{q},\lambda^{\prime}}en_{0}\delta(z) & n_{0}m\left(\omega_{\mathbf{q},\lambda^{\prime}}^{2}-\beta^{2}\mathbf{q}\otimes\mathbf{q}\right)\delta(z)
\end{array}\right]\left[\begin{array}{c}
\mathbf{A}_{\mathbf{q},\lambda^{\prime}}(z)\\
\bm{\upsilon}_{\mathbf{q},\lambda^{\prime}}
\end{array}\right]=0.\label{eq:mode2}
\end{equation}
Now we contract Eq.~\eqref{eq:mode1} with $\left[\begin{array}{cc}
\mathbf{A}_{\mathbf{q},\lambda^{\prime}}^{\dagger}(z) & \bm{\upsilon}_{\mathbf{q},\lambda^{\prime}}^{\dagger}\end{array}\right]$, Eq.~\eqref{eq:mode1} with $\left[\begin{array}{cc}
\mathbf{A}_{\mathbf{q},\lambda}^{\dagger}(z) & \bm{\upsilon}_{\mathbf{q},\lambda}^{\dagger}\end{array}\right]$ and integrate both equations over $z$ obtaining
\begin{align}
\int dz\left[\begin{array}{cc}
\mathbf{A}_{\mathbf{q},\lambda^{\prime}}^{\dagger}(z) & \bm{\upsilon}_{\mathbf{q},\lambda^{\prime}}^{\dagger}\end{array}\right]\left[\begin{array}{cc}
\omega_{\mathbf{q},\lambda}^{2}\epsilon_{0}\bar{\epsilon}_{d}(z)-\frac{1}{\mu_{0}}D_{\mathbf{q}}\times D_{\mathbf{q}}\times & i\omega_{\mathbf{q},\lambda}en_{0}\delta(z)\\
-i\omega_{\mathbf{q},\lambda}en_{0}\delta(z) & n_{0}m\left(\omega_{\mathbf{q},\lambda}^{2}-\beta^{2}\mathbf{q}\otimes\mathbf{q}\right)\delta(z)
\end{array}\right]\left[\begin{array}{c}
\mathbf{A}_{\mathbf{q},\lambda}(z)\\
\bm{\upsilon}_{\mathbf{q},\lambda}
\end{array}\right] & =0,\label{eq:ortho1mid1}\\
\int dz\left[\begin{array}{cc}
\mathbf{A}_{\mathbf{q},\lambda}^{\dagger}(z) & \bm{\upsilon}_{\mathbf{q},\lambda}^{\dagger}\end{array}\right]\left[\begin{array}{cc}
\omega_{\mathbf{q},\lambda^{\prime}}^{2}\epsilon_{0}\bar{\epsilon}_{d}(z)-\frac{1}{\mu_{0}}D_{\mathbf{q}}\times D_{\mathbf{q}}\times & i\omega_{\mathbf{q},\lambda^{\prime}}en_{0}\delta(z)\\
-i\omega_{\mathbf{q},\lambda^{\prime}}en_{0}\delta(z) & n_{0}m\left(\omega_{\mathbf{q},\lambda^{\prime}}^{2}-\beta^{2}\mathbf{q}\otimes\mathbf{q}\right)\delta(z)
\end{array}\right]\left[\begin{array}{c}
\mathbf{A}_{\mathbf{q},\lambda^{\prime}}(z)\\
\bm{\upsilon}_{\mathbf{q},\lambda^{\prime}}
\end{array}\right] & =0.\label{eq:ortho1mid2}
\end{align}
Taking the conjugate of Eq.~\eqref{eq:ortho1mid2}, we obtain
\begin{equation}
\int dz\left[\begin{array}{cc}
\mathbf{A}_{\mathbf{q},\lambda^{\prime}}^{\dagger}(z) & \bm{\upsilon}_{\mathbf{q},\lambda^{\prime}}^{\dagger}\end{array}\right]\left[\begin{array}{cc}
\omega_{\mathbf{q},\lambda^{\prime}}^{2}\epsilon_{0}\bar{\epsilon}_{d}(z)-\frac{1}{\mu_{0}}D_{\mathbf{q}}\times D_{\mathbf{q}}\times & i\omega_{\mathbf{q},\lambda^{\prime}}en_{0}\delta(z)\\
-i\omega_{\mathbf{q},\lambda^{\prime}}en_{0}\delta(z) & n_{0}m\left(\omega_{\mathbf{q},\lambda^{\prime}}^{2}-\beta^{2}\mathbf{q}\otimes\mathbf{q}\right)\delta(z)
\end{array}\right]\left[\begin{array}{c}
\mathbf{A}_{\mathbf{q},\lambda}(z)\\
\bm{\upsilon}_{\mathbf{q},\lambda}
\end{array}\right]=0.
\end{equation}
Subtracting this last equation from Eq.~\eqref{eq:ortho1mid1}, we
obtain
\begin{equation}
\int dz\left[\begin{array}{cc}
\mathbf{A}_{\mathbf{q},\lambda^{\prime}}^{\dagger}(z) & \bm{\upsilon}_{\mathbf{q},\lambda^{\prime}}^{\dagger}\end{array}\right]\left[\begin{array}{cc}
\left(\omega_{\mathbf{q},\lambda}^{2}-\omega_{\mathbf{q},\lambda^{\prime}}^{2}\right)\epsilon_{0}\bar{\epsilon}_{d}(z) & i\left(\omega_{\mathbf{q},\lambda}-\omega_{\mathbf{q},\lambda^{\prime}}\right)en_{0}\delta(z)\\
-i\left(\omega_{\mathbf{q},\lambda}-\omega_{\mathbf{q},\lambda^{\prime}}\right)en_{0}\delta(z) & n_{0}m\left(\omega_{\mathbf{q},\lambda}^{2}-\omega_{\mathbf{q},\lambda^{\prime}}^{2}\right)\delta(z)
\end{array}\right]\left[\begin{array}{c}
\mathbf{A}_{\mathbf{q},\lambda}(z)\\
\bm{\upsilon}_{\mathbf{q},\lambda}
\end{array}\right]=0.
\end{equation}
If $\omega_{\mathbf{q},\lambda}\neq\omega_{\mathbf{q},\lambda^{\prime}}$,
we can divide by $\omega_{\mathbf{q},\lambda}-\omega_{\mathbf{q},\lambda^{\prime}}$
obtaining one of the orthogonality conditions:
\begin{equation}
\int dz\left[\begin{array}{cc}
\mathbf{A}_{\mathbf{q},\lambda^{\prime}}^{\dagger}(z) & \bm{\upsilon}_{\mathbf{q},\lambda^{\prime}}^{\dagger}\end{array}\right]\left[\begin{array}{cc}
\left(\omega_{\mathbf{q},\lambda}+\omega_{\mathbf{q},\lambda^{\prime}}\right)\epsilon_{0}\bar{\epsilon}_{d}(z) & ien_{0}\delta(z)\\
-ien_{0}\delta(z) & n_{0}m\left(\omega_{\mathbf{q},\lambda}-\omega_{\mathbf{q},\lambda^{\prime}}\right)\delta(z)
\end{array}\right]\left[\begin{array}{c}
\mathbf{A}_{\mathbf{q},\lambda}(z)\\
\bm{\upsilon}_{\mathbf{q},\lambda}
\end{array}\right]=0.\label{eq:orthogonality_1}
\end{equation}

We can obtain an additional orthogonality relation. If we take the
complex conjugate of Eq.~\eqref{eq:ortho1mid1}, replace $\mathbf{q}\rightarrow-\mathbf{q}$
and $\lambda\rightarrow\lambda^{\prime}$, and using the fact that
$\omega_{\mathbf{q},\lambda}=\omega_{-\mathbf{q},\lambda}$, we obtain
\begin{equation}
\left[\begin{array}{cc}
\omega_{\mathbf{q},\lambda^{\prime}}^{2}\epsilon_{0}\bar{\epsilon}_{d}(z)-\frac{1}{\mu_{0}}D_{\mathbf{q}}\times D_{\mathbf{q}}\times & -i\omega_{\mathbf{q},\lambda^{\prime}}en_{0}\delta(z)\\
i\omega_{\mathbf{q},\lambda^{\prime}}en_{0}\delta(z) & n_{0}m\left(\omega_{\mathbf{q},\lambda^{\prime}}^{2}-\beta^{2}\mathbf{q}\otimes\mathbf{q}\right)\delta(z)
\end{array}\right]\left[\begin{array}{c}
\mathbf{A}_{-\mathbf{q},\lambda}^{*}(z)\\
\bm{\upsilon}_{-\mathbf{q},\lambda}^{*}
\end{array}\right]=0.\label{eq:mode3}
\end{equation}
We now contract this equation with $\left[\begin{array}{cc}
\mathbf{A}_{\mathbf{q},\lambda}^{\dagger}(z) & \bm{\upsilon}_{\mathbf{q},\lambda}^{\dagger}\end{array}\right]$, integrate over $z$ and take its complex conjugate, obtaining
\begin{equation}
\int dz\left[\begin{array}{cc}
\mathbf{A}_{-\mathbf{q},\lambda}^{t}(z) & \bm{\upsilon}_{-\mathbf{q},\lambda}^{t}\end{array}\right]\left[\begin{array}{cc}
\omega_{\mathbf{q},\lambda^{\prime}}^{2}\epsilon_{0}\bar{\epsilon}_{d}(z)-\frac{1}{\mu_{0}}D_{\mathbf{q}}\times D_{\mathbf{q}}\times & -i\omega_{\mathbf{q},\lambda^{\prime}}en_{0}\delta(z)\\
i\omega_{\mathbf{q},\lambda^{\prime}}en_{0}\delta(z) & n_{0}m\left(\omega_{\mathbf{q},\lambda^{\prime}}^{2}-\beta^{2}\mathbf{q}\otimes\mathbf{q}\right)\delta(z)
\end{array}\right]\left[\begin{array}{c}
\mathbf{A}_{\mathbf{q},\lambda}(z)\\
\bm{\upsilon}_{\mathbf{q},\lambda}
\end{array}\right]=0.\label{eq:ortho2_mid1}
\end{equation}
Contracting Eq.~\eqref{eq:mode1} with $\left[\begin{array}{cc}
\mathbf{A}_{-\mathbf{q},\lambda}^{t}(z) & \bm{\upsilon}_{-\mathbf{q},\lambda}^{t}\end{array}\right]$ and integrating over $z$ we obtain
\begin{equation}
\int dz\left[\begin{array}{cc}
\mathbf{A}_{-\mathbf{q},\lambda}^{t}(z) & \bm{\upsilon}_{-\mathbf{q},\lambda}^{t}\end{array}\right]\left[\begin{array}{cc}
\omega_{\mathbf{q},\lambda}^{2}\epsilon_{0}\bar{\epsilon}_{d}(z)-\frac{1}{\mu_{0}}D_{\mathbf{q}}\times D_{\mathbf{q}}\times & i\omega_{\mathbf{q},\lambda}en_{0}\delta(z)\\
-i\omega_{\mathbf{q},\lambda}en_{0}\delta(z) & n_{0}m\left(\omega_{\mathbf{q},\lambda}^{2}-\beta^{2}\mathbf{q}\otimes\mathbf{q}\right)\delta(z)
\end{array}\right]\left[\begin{array}{c}
\mathbf{A}_{\mathbf{q},\lambda}(z)\\
\bm{\upsilon}_{\mathbf{q},\lambda}
\end{array}\right]=0.\label{eq:ortho2_mid2}
\end{equation}
Subtracting Eq.~\eqref{eq:ortho2_mid1} from \eqref{eq:ortho2_mid2},
we obtain
\begin{equation}
\int dz\left[\begin{array}{cc}
\mathbf{A}_{-\mathbf{q},\lambda}^{t}(z) & \bm{\upsilon}_{-\mathbf{q},\lambda}^{t}\end{array}\right]\left[\begin{array}{cc}
\left(\omega_{\mathbf{q},\lambda}^{2}-\omega_{\mathbf{q},\lambda^{\prime}}^{2}\right)\epsilon_{0}\bar{\epsilon}_{d}(z) & i\left(\omega_{\mathbf{q},\lambda}+\omega_{\mathbf{q},\lambda^{\prime}}\right)en_{0}\delta(z)\\
-i\left(\omega_{\mathbf{q},\lambda}+\omega_{\mathbf{q},\lambda^{\prime}}\right)en_{0}\delta(z) & n_{0}m\left(\omega_{\mathbf{q},\lambda}^{2}-\omega_{\mathbf{q},\lambda^{\prime}}^{2}\right)\delta(z)
\end{array}\right]\left[\begin{array}{c}
\mathbf{A}_{\mathbf{q},\lambda}(z)\\
\bm{\upsilon}_{\mathbf{q},\lambda}
\end{array}\right]=0.
\end{equation}
For $\omega_{\mathbf{q},\lambda}+\omega_{\mathbf{q},\lambda^{\prime}}\neq0$,
we obtain a second orthogonality condition:
\begin{equation}
\int dz\left[\begin{array}{cc}
\mathbf{A}_{-\mathbf{q},\lambda}^{t}(z) & \bm{\upsilon}_{-\mathbf{q},\lambda}^{t}\end{array}\right]\left[\begin{array}{cc}
\left(\omega_{\mathbf{q},\lambda}-\omega_{\mathbf{q},\lambda^{\prime}}\right)\epsilon_{0}\bar{\epsilon}_{d}(z) & ien_{0}\delta(z)\\
-ien_{0}\delta(z) & n_{0}m\left(\omega_{\mathbf{q},\lambda}-\omega_{\mathbf{q},\lambda^{\prime}}\right)\delta(z)
\end{array}\right]\left[\begin{array}{c}
\mathbf{A}_{\mathbf{q},\lambda}(z)\\
\bm{\upsilon}_{\mathbf{q},\lambda}
\end{array}\right]=0.\label{eq:orthogonality_2}
\end{equation}

\subsection{Hamiltonian in term of mode amplitudes}

We now insert the expansion of the fields in terms of mode-functions,
Eqs.~\eqref{eq:mode_expansion_A} and \eqref{eq:mode_expansion_v},
into the classical Hamiltonian \eqref{eq:Hamiltonian_hydro_classical}.
Using Eqs.~\eqref{eq:conjugate_A} and \eqref{eq:conjugate_v}, we
can write
\begin{equation}
H=\frac{1}{2}\sum_{\mathbf{q},\lambda\lambda^{\prime}}\left(h_{\mathbf{\lambda,\lambda^{\prime}}}(\mathbf{q})\alpha_{\mathbf{q},\lambda^{\prime}}^{*}(t)\alpha_{\mathbf{q},\lambda}(t)+\tilde{h}_{\mathbf{\lambda,\lambda^{\prime}}}(\mathbf{q})\alpha_{-\mathbf{q},\lambda^{\prime}}(t)\alpha_{\mathbf{q},\lambda}(t)\right)+\text{c.c.},
\end{equation}
where
\begin{align}
h_{\mathbf{\lambda,\lambda^{\prime}}}(\mathbf{q}) & =\int dz\left[\begin{array}{cc}
\mathbf{A}_{\mathbf{q},\lambda^{\prime}}^{\dagger}(z) & \bm{\upsilon}_{\mathbf{q},\lambda^{\prime}}^{\dagger}\end{array}\right]\nonumber \\
 & \left[\begin{array}{cc}
\omega_{\mathbf{q},\lambda^{\prime}}\omega_{\mathbf{q},\lambda}\epsilon_{0}\bar{\epsilon}_{d}(z)+\mu_{0}^{-1}D_{\mathbf{q}}\times D_{\mathbf{q}} & 0\\
0 & \delta(z)\left(n_{0}m\omega_{\mathbf{q},\lambda^{\prime}}\omega_{\mathbf{q},\lambda}+n_{0}m\beta^{2}\mathbf{q}\otimes\mathbf{q}\right)
\end{array}\right]\left[\begin{array}{c}
\mathbf{A}_{\mathbf{q},\lambda}(z)\\
\bm{\upsilon}_{\mathbf{q},\lambda}
\end{array}\right]\\
\tilde{h}_{\mathbf{\lambda,\lambda^{\prime}}}(\mathbf{q}) & =\int dz\left[\begin{array}{cc}
\mathbf{A}_{-\mathbf{q},\lambda^{\prime}}^{t}(z) & \bm{\upsilon}_{-\mathbf{q},\lambda^{\prime}}^{t}\end{array}\right]\nonumber \\
 & \left[\begin{array}{cc}
-\omega_{\mathbf{q},\lambda^{\prime}}\omega_{\mathbf{q},\lambda}\epsilon_{0}\bar{\epsilon}_{d}(z)+\mu_{0}^{-1}D_{\mathbf{q}}\times D_{\mathbf{q}} & 0\\
0 & \delta(z)\left(-n_{0}m\omega_{\mathbf{q},\lambda^{\prime}}\omega_{\mathbf{q},\lambda}+n_{0}m\beta^{2}\mathbf{q}\otimes\mathbf{q}\right)
\end{array}\right]\left[\begin{array}{c}
\mathbf{A}_{\mathbf{q},\lambda}(z)\\
\bm{\upsilon}_{\mathbf{q},\lambda}
\end{array}\right]
\end{align}
Using the mode-function equation~\eqref{eq:mode1}, we can write
\begin{align}
h_{\mathbf{\lambda,\lambda^{\prime}}}(\mathbf{q}) & =\omega_{\mathbf{q},\lambda}\int dz\left[\begin{array}{cc}
\mathbf{A}_{\mathbf{q},\lambda^{\prime}}^{\dagger}(z) & \bm{\upsilon}_{\mathbf{q},\lambda^{\prime}}^{\dagger}\end{array}\right]\left[\begin{array}{cc}
\left(\omega_{\mathbf{q},\lambda^{\prime}}+\omega_{\mathbf{q},\lambda}\right)\epsilon_{0}\bar{\epsilon}_{d}(z) & ien_{0}\delta(z)\\
-ien_{0}\delta(z) & \delta(z)n_{0}m\left(\omega_{\mathbf{q},\lambda^{\prime}}+\omega_{\mathbf{q},\lambda}\right)
\end{array}\right]\left[\begin{array}{c}
\mathbf{A}_{\mathbf{q},\lambda}(z)\\
\bm{\upsilon}_{\mathbf{q},\lambda}
\end{array}\right],\\
\tilde{h}_{\mathbf{\lambda,\lambda^{\prime}}}(\mathbf{q}) & =\omega_{\mathbf{q},\lambda}\int dz\left[\begin{array}{cc}
\mathbf{A}_{-\mathbf{q},\lambda^{\prime}}^{t}(z) & \bm{\upsilon}_{-\mathbf{q},\lambda^{\prime}}^{t}\end{array}\right]\left[\begin{array}{cc}
\left(\omega_{\mathbf{q},\lambda}-\omega_{\mathbf{q},\lambda^{\prime}}\right)\epsilon_{0}\bar{\epsilon}_{d}(z) & ien_{0}\delta(z)\\
-ien_{0}\delta(z) & \delta(z)n_{0}m\left(\omega_{\mathbf{q},\lambda}-\omega_{\mathbf{q},\lambda^{\prime}}\right)
\end{array}\right]\left[\begin{array}{c}
\mathbf{A}_{\mathbf{q},\lambda}(z)\\
\bm{\upsilon}_{\mathbf{q},\lambda}
\end{array}\right].
\end{align}
Using the orthogonality condition \eqref{eq:orthogonality_2} we conclude
that $\tilde{h}_{\mathbf{\lambda,\lambda^{\prime}}}(\mathbf{q})=0$.
The orthogonality condition \eqref{eq:orthogonality_1} implies that
$h_{\mathbf{\lambda,\lambda^{\prime}}}(\mathbf{q})=0$, except when
$\lambda=\lambda^{\prime}$ (assuming there are no degeneracies in
$\omega_{\mathbf{q},\lambda}$). Therefore, we see that the classical
Hamiltonian can be written as in Eq.~\eqref{eq:classical_hamiltonian_mode},
with the mode-length being biven by
\begin{equation}
L_{\mathbf{q},\lambda}=\frac{1}{2\epsilon_{0}\omega_{\mathbf{q},\lambda}^{2}}\int dz\left[\begin{array}{cc}
\mathbf{A}_{\mathbf{q},\lambda^{\prime}}^{\dagger}(z) & \bm{\upsilon}_{\mathbf{q},\lambda^{\prime}}^{\dagger}\end{array}\right]\left[\begin{array}{cc}
2\omega_{\mathbf{q},\lambda}^{2}\epsilon_{0}\bar{\epsilon}_{d}(z) & i\omega_{\mathbf{q},\lambda}en_{0}\delta(z)\\
-i\omega_{\mathbf{q},\lambda}en_{0}\delta(z) & 2\omega_{\mathbf{q},\lambda}^{2}\delta(z)n_{0}m
\end{array}\right]\left[\begin{array}{c}
\mathbf{A}_{\mathbf{q},\lambda}(z)\\
\bm{\upsilon}_{\mathbf{q},\lambda}
\end{array}\right],
\end{equation}
which can be written as Eq.~\eqref{eq:hydro_mode_length}.

\end{widetext}

\end{document}